\documentclass[12pt]{article}
\usepackage{jheppub}
\usepackage{amssymb,amsmath}
\usepackage{latexsym}
\usepackage{graphicx}
\usepackage{booktabs}
\usepackage{bbm}
\usepackage{color}

\def\be{\begin{equation}}
\def\ee{\end{equation}}
\def\amp{\mathfrak{a}}
\def\Kt{\mathcal{K}}
\def\Kn{\overline{\mathcal{K}}}
\def\sd{\mathfrak{s}}
\def\HnG#1{\left(\mathfrak{h}_{_2(#1)}\right)}
\def\HnL#1{\left(\mathfrak{h}_{_m(#1)}\right)}
\def\Hntot#1{\left(\mathfrak{h}_{_\text{sum}(#1)}\right)}

\definecolor{rust}{rgb}{0.8,0.2,0.2}
\definecolor{purple}{rgb}{0.8,0.1,0.9}
\definecolor{olivegreen}{rgb}{0,0.52,0.17}

\title{Towards a second law for Lovelock theories}

\author[a]{Sayantani Bhattacharyya}
\author[b]{\!, Felix M. Haehl}
\author[c]{\!, Nilay Kundu}
\author[d]{\!, R.\ Loganayagam}
\author[e]{\!, Mukund Rangamani}
\affiliation[\,a]{Indian Institute of Technology Kanpur, Kanpur 208016,  India.}
\affiliation[\,b]{Department of Physics and Astronomy, University of British Columbia,\\
6224 Agricultural Road, Vancouver, B.C.\ V6T 1Z1, Canada.}
\affiliation[\,c]{Center for Gravitational Physics, Yukawa Institute for Theoretical Physics (YITP), \\ Kyoto University, Kyoto 606-8502, Japan}
\affiliation[\,d]{International Centre for Theoretical Sciences (ICTS-TIFR), \\
Shivakote, Hesaraghatta Hobli, Bengaluru 560089, India.}
\affiliation[\,e]{
Center for Quantum Mathematics and Physics (QMAP),  \\
Department of Physics, University of California, Davis, CA 95616 USA.}
\emailAdd{sayanta@iitk.ac.in}
\emailAdd{f.m.haehl@gmail.com}
\emailAdd{nilay.tifr@gmail.com}
\emailAdd{nayagam@gmail.com}
\emailAdd{mukund@physics.ucdavis.edu}

\abstract{In classical general relativity described by Einstein-Hilbert gravity, black holes behave as thermodynamic objects. In particular, the laws of black hole mechanics can be interpreted as laws of thermodynamics. The first law of black hole mechanics extends to higher derivative theories via the Noether charge construction of Wald. One also expects the statement of the second law, which in Einstein-Hilbert theory owes to Hawking's area theorem, to extend to higher derivative theories. To argue for this however one needs a notion of entropy for dynamical black holes, which the Noether charge construction does not provide. We propose such an entropy function for the family of Lovelock theories, treating the higher derivative terms as perturbations to the Einstein-Hilbert theory. Working around a dynamical black hole solution, and making no assumptions about the amplitude of departure from equilibrium, we construct a candidate entropy functional valid to all orders in the  low energy effective field theory. This entropy functional satisfies a second law, modulo a certain subtle boundary term, which deserves further investigation in non-spherically symmetric situations.}

\begin{document}

\begin{flushright} 
\small{YITP-16-135} 
\end{flushright}

\maketitle
\flushbottom

\section{Introduction}
\label{sec:intro}

On general grounds, upon taking the low energy limit of any UV complete quantum theory of gravity, we expect to generate higher derivative corrections to the leading two-derivative Einstein-Hilbert action with matter:
\begin{equation}
I= \frac{1}{16\pi\, G_N^{(d)}}\
\int d^dx \sqrt{-g}~ \left( R + \mathcal{L}_{matter} \right)  \,.
\label{einhilactn0}
\end{equation}
The precise form of the higher derivative corrections depend on the nature of the UV completion. While it is expected that string theory provides the desired UV complete description, we do not yet know how these are organized. Therefore, it is in principle useful, if using general principles, we could constrain the  space of admissible low energy theories. This strategy has previously been employed with some success both in non-gravitational \cite{Adams:2006sv}, and gravitational \cite{Camanho:2014apa} settings.

The second law of thermodynamics is one such principle that we can test on low energy solutions of gravity. As is well known, black holes are thermodynamic objects \cite{Hawking:1971tu, Bardeen:1973gs,  Bekenstein:1973ur, Hawking:1974sw}. We can associate extensive parameters, such as energy, and other conserved charges, in terms of asymptotic data, while intensive parameters
such as temperature and chemical potential refer to properties of the horizon.
In the Einstein-Hilbert theory Eq.~\eqref{einhilactn0} we associate a notion of entropy, given by the area of the event horizon. Thanks to Hawking's famous area theorem, which states that the horizon area is non-decreasing in any physical process (for  matter  satisfying suitable energy conditions), this notion of entropy satisfies the second law of thermodynamics (see e.g., \cite{Hawking:1973uf}).

A natural question then is how does this picture extend to higher derivative theories. This question is especially salient since a study of higher derivative black hole dynamics in the non-linear regime
could help us constrain such terms via gravitational wave observations. Thus, the question of whether  higher derivative black holes are sensible endpoints of dynamics  in a second law sense
is an interesting question to answer. The question of extending black hole thermodynamics to higher derivative gravity  was first tackled by various people more than two decades ago,
and culminated in a beautiful application of the Noether procedure in the covariant phase space formalism  \cite{Wald:1993nt, Iyer:1994ys}. In particular,
Wald argued that for stationary black hole solutions of higher derivative gravity theories, the entropy is a Noether charge associated with time translations along the horizon generating Killing field. This Wald entropy was constructed to explicitly satisfy the first law of thermodynamics, which being an equilibrium statement, can be understood in the stationary solution.

An open question since then has been whether there is a notion of second law of black hole mechanics in higher derivative gravity. This question was addressed in various guises over the years,  \cite{Jacobson:1993xs,Jacobson:1993vj,Iyer:1994ys, Jacobson:1995uq, Kolekar:2012tq, Sarkar:2013swa, Bhattacharjee:2015yaa, Wall:2015raa, Bhattacharjee:2015qaa}. We will say more about these works below, but first let us appreciate the issues involved.

The second law involves dynamics. In its crudest version it says that, in a physical process in which a system evolves from one equilibrium configuration to another, the total entropy must increase. Whilst this is a non-local statement, comparing only the initial and final configurations, one can ensure this by simply exhibiting  a function of the system variables, the \emph{entropy function},
which is monotone under time evolution, and reduces in equilibrium to the familiar notion of entropy. These requirements, per se, seem quite unrestrictive, for we make no demand of uniqueness.

Constructing an entropy function in a higher derivative theory will convince us of the validity of the second law. On the contrary, if we can show that no such entropy function exists, then we would be forced to conclude that the higher derivative theory is physically unacceptable.\footnote{ We do not wish to suggest that the existence of the entropy function is the most stringent condition one could impose. Demanding causality and unitarity may imply stronger constraints, which may rule out a vaster swathe of the higher derivative landscape. It is nevertheless interesting to ask if the entropy function itself serves to give us interesting constraints on the admissible higher derivative terms.} All told, we wish to basically examine whether dynamical black holes in higher derivative theories
can indeed be viewed as thermodynamic objects.

In gravity, a  configuration in thermal equilibrium, translates to geometry with a Killing horizon. Recall that the event horizon is a null surface, ruled by null geodesics. The equilibrium configuration is one where the null generators are along a Killing direction. In Einstein-Hilbert theory Eq.~\eqref{einhilactn0}, the area of the horizon provides a suitable entropy function, thanks to the aforementioned  area increase theorem.  Assuming the existence of a Killing horizon, Wald derived
the Wald entropy function \cite{Wald:1993nt,Iyer:1994ys}, which gives us a notion of equilibrium entropy in higher derivative theories. We then need to find its extension to the dynamical setting, satisfying the desired monotonicity conditions.

Let us try to organize the discussion in a useful manner. Firstly, higher derivative terms come with characteristic length scales (to soak up dimensions). Let us assume that these are all calibrated against a single `string' scale $\ell_s$, so that we have a bunch of dimensionless couplings, collectively denoted as $\alpha$, which parametrize the higher derivative theory. We pick a stationary black hole in this theory, which will determine the initial equilibrium configuration. To study the second law, we should now be perturbing away from equilibrium, which itself can be parametrized by a) the amplitude, $\amp$, of the departure from equilibrium, and b) the characteristic frequency, $\omega$, of the disturbance.\footnote{ We assume for the moment that the frequency scale is commensurate with spatial momenta in the case of non-compactly generated horizons.}
A putative entropy function should, at  the very least, carry information about these three parameters. The following summarizes what is known to date:
\begin{itemize}
\item The Wald entropy can be constructed for arbitrary $\alpha$ with $\amp=0$.
\item In \cite{Jacobson:1993xs,Jacobson:1995uq} the authors construct entropy functions for theories where the higher derivative Lagrangian is written solely in terms of the Ricci scalar (the so-called $f(R)$ theories), for a finite range of $\alpha$, and arbitrary $\amp$ and $\omega$.
\item In recent years,  \cite{Kolekar:2012tq} first constructed an entropy functional that was valid for higher derivative interactions of the Lovelock form (see below) for small amplitude departures from equilibrium $\amp\ll 1$. This was generalized subsequently to $f(\text{Lovelock})$ theories in \cite{Sarkar:2013swa}.
\item These constructions were further examined  in \cite{Bhattacharjee:2015yaa} who showed that by fixing various ambiguities associated with  foliations of the event horizon, one could construct an entropy function for four-derivative theories of gravity (in spherically symmetric situations). Again the range of validity was $\amp \ll 1$.
\item In a more interesting development, \cite{Wall:2015raa} showed that for any higher derivative theory of gravity, a particular correction to the Wald functional due to \cite{Fursaev:2013fta,Dong:2013qoa,Camps:2013zua}, which construct functions for computing entanglement entropy in the holographic context, provides an entropy function to linear order in amplitudes  $\amp \ll 1$.
\item While, there is no explicit discussion in the literature, the explicit  construction of an entropy current  \cite{Bhattacharyya:2008xc}  in the fluid/gravity correspondence \cite{Bhattacharyya:2008jc,Hubeny:2011hd}, is strongly
 suggestive of an entropy function in higher derivative gravity perturbatively in the couplings and the frequency, $\alpha \ll 1, \,\omega \ell_s \ll1$, but valid for arbitrary amplitudes.\footnote{ In the hydrodynamic regime, the transport coefficients become unphysical, e.g., the shear viscosity goes negative, for finite values of the higher derivative coupling \cite{Brigante:2007nu}. A negative value of viscosity leads to entropy destruction, but this constraint, one must remark, is far weaker than demanding that the theory respect causality, either in the effective field theory sense \cite{Brigante:2007nu}, or from a fundamental perspective \cite{Camanho:2014apa}.}
\end{itemize}

In this note, we shall construct entropy function for the Lovelock family of higher derivative Lagrangians. We will work perturbatively in higher derivative interactions, treating, as one should, the corrections to Einstein-Hilbert theory in a gradient expansion. 
The effective small parameter governing our perturbation theory is going to be 
$\omega\, \ell_s$ which is dimensionless. We however will make no assumption about the amplitude $\amp$  of the  perturbation away from equilibrium. In fact,  we allow arbitrary time evolution away from equilibrium; our only proviso is that this time evolution be sensibly captured by the low energy effective field theory.  In geometric terms, we will allow fluctuations of the black hole horizon which, at the horizon scale, are small enough, so that the leading Einstein-Hilbert term dominates over the higher derivative terms.

Despite the two-derivative theory dominating on the horizon scale, we will need to modify the  entropy  to respect the second law. This follows because, while the leading area contribution is per se large, it remains possible that under evolution, the area variation is anomalously small. This can then be overwhelmed by the higher derivative 
$\mathcal{O}(\omega \ell_s)$ contributions, spoiling the monotonicity of the entropy. To ensure that this does not happen we will need to shift the entropy function away from the Wald form with suitable corrections.

 The outline of this paper is as follows. We begin in \S\ref{sec:state} with a basic statement of the problem we study, and a complete summary of our results. We will then describe in \S\ref{sec:GaussBonnet} a construction of an all-order entropy function for Gauss-Bonnet theory. Our analysis will give an entropy function with monotonicity properties in spherically symmetric configurations, but we note an interesting obstruction (in the form of a total derivative), which precludes us form making a general statement. We elaborate on this in the course of our discussion. We then show in \S\ref{sec:lovelock} how to extend the construction to higher Lovelock terms, once again encountering a total derivative term. The final result for an arbitrary combination of Lovelock terms can be compactly packaged in terms of a variations of the gravitational Lagrangian with respect to the curvatures, see Eq.~\eqref{eq:SfinalIntro}. In general, as long as the total derivative obstruction term\footnote{ This term also can be expressed in terms of the variation of the Lagrangian with respect to curvatures, multiplying a certain tensor built out of the extrinsic curvatures, cf., Eq.~\eqref{eq:Xobsgen}.} is negative definite (or vanishes as in spherical symmetry), we have our desired entropy function. Finally, in \S\ref{sec:discussion} we comment on potential generalizations and open issues. The various appendices collect some useful technical information relating to the computations. 

\subsection{Statement of the problem}
\label{sec:state}
Let us first define the problem we tackle more precisely. We consider higher derivative corrections to Eq.~\eqref{einhilactn0}. Schematically, suppressing all indices,  these corrections will be of the form $\alpha_k\,\ell_s^k \, D^k \, R$. $\ell_s$ here is the length scale at which higher derivative corrections appear. Rather than focusing on all possible corrections of this form, we will restrict attention to the family of Lovelock theories, whose action can be written as:\footnote{ To keep the expressions readable we have chosen to work in natural entropy units by setting $G_N^{(d)} = \frac{1}{4}$. Thus an overall dimensionful dependence on the Planck scale is suppressed in our analysis, which however, can be restored easily if desired.}
\begin{equation}
\begin{split}
I &= \frac{1}{4\pi}\ \int d^dx\,  \sqrt{-g} \left(R + \sum_{m=2}^\infty \, \alpha_m\, 
\ell_s^{2m-2} \, \mathcal{L}_m + \mathcal{L}_{matter} \right)\\
\mathcal{L}_m &=
\delta^{\mu_1\nu_1\cdots \mu_m \nu_m}_{\rho_1\sigma_1\cdots \rho_m \sigma_m} \  R^{\rho_1}{}_{\mu_1}{}^{\sigma_1}{}_{\nu_1} \ \cdots\  R^{\rho_m}{}_{\mu_m}{}^{\sigma_m}{}_{\nu_m} \,.
\end{split}
\label{eq:action1}
\end{equation}
We have written the Lovelock action in terms of the generalized Kronecker symbol
$\delta_{\rho_1 \sigma_1\cdots \rho_m \sigma_m}^{\mu_1 \nu_1\cdots \mu_m \nu_m}$.\footnote{ $\delta^{\mu_1\mu_2 \cdots \mu_m}_{\nu_1\nu_2\cdots \nu_m}$ is the determinant of an $m\times m$ matrix  whose $(ij)^{\rm th}$  element is given by $\delta^{\mu_i}_{\nu_j}$.
Hence $\delta_{\rho_1 \sigma_1\cdots \rho_m \sigma_m}^{\mu_1 \nu_1\cdots \mu_m \nu_m}$ is completely antisymmetric both in all its upper and lower indices.}
 In particular, the first non-trivial contribution at the four-derivative order is given by the Gauss-Bonnet theory
\begin{equation}
\mathcal{L}_2 \equiv \mathcal{L}_{GB} =
R^2 - 4 R_{\mu\nu} R^{\mu\nu} +R_{\mu\nu\alpha\beta} R^{\mu\nu\alpha\beta} \,,
\label{eq:gb}
\end{equation}

The matter Lagrangian  $\mathcal{L}_{matter}$ will play a peripheral role in our analysis. The only assumption we make is that the matter we include satisfies the null energy condition (NEC), $T^{matter}_{\mu\nu} \,k^\mu k^\nu \geq 0$ for any null $k^\mu$.
For the initial part of the paper we will explain the construction in the case of the Gauss-Bonnet theory \S\ref{sec:GaussBonnet}, and only thence generalize to the Lovelock case \S\ref{sec:lovelock}.
 
As we are interested in analyzing black hole geometries we will assume that we have been handed a solution to a spacetime metric $g_{\mu\nu}$ which has a regular horizon. We pick a coordinate chart for this geometry which manifests regularity at the horizon (similar to the ingoing Eddington-Finkelstein chart), and parameterize the metric as\footnote{ We thank Shiraz Minwalla for numerous discussions on the structure of the entropy function, and setting up useful coordinate charts to construct candidate functions.}
\begin{equation}
\begin{split}
&ds^2  = 2\, dv~ dr - f(r,v, {\bf x})\, dv^2 + 2 \, k_A (r,v, {\bf x}) \, dv~ dx^A + h_{AB}(r,v, {\bf x})\,dx^A dx^B\,.
\end{split}
\label{metricintro}
\end{equation}
The null hypersurface of the horizon $\mathcal{H}^+$ is the locus $r=0$ where several of these functions vanish, e.g., 
\begin{equation}\label{metricintroa}
\begin{split}
f(r,v, {\bf x})\big|_{\mathcal{H}^+}=  k_A (r,v, {\bf x})\big|_{\mathcal{H}^+} = \partial_r f(r,v, {\bf x})\big|_{\mathcal{H}^+} =0\,.
\end{split}
\end{equation}
In \S\ref{sec:geom} we shall see that such a coordinate choice is always possible if the spacetime contains a null hypersurface. From Eq.~\eqref{metricintro} it follows that the horizon is at $r=0$, with $\partial_v$ being the affinely parametrized null generators of the horizon. This choice is quite natural from the fluid/gravity intuition (it was also previously employed in \cite{Wall:2015raa}). Spatial sections of the horizon at constant $v$ slices are denoted as $\Sigma_v$.

Our goal is to construct an entropy function, $S_{\text{total}}$, in terms of the derivatives of metric components evaluated on the horizon,  subject to the following conditions:
\begin{itemize}
\item \label{cond1} {Condition 1 {\bf (C1)}:}  $\partial_v S_{\text{total}} \ge 0$:
on every black hole solution captured by metric of the form Eq.~\eqref{metricintro}, and for every finite value of $v$.
\item{Condition 2 {\bf (C2)}:} \label{cond2}$S_{\text{total}}$ reduces to Wald entropy $S_\text{Wald}$ in equilibrium (whence $\partial_v$ is a Killing vector).
\end{itemize}

As indicated our analysis will be carried out in a gradient expansion, with the perturbation parameter being $\omega \,\ell_s$, consistent with the rules of effective field theory.  $\ell_s$ is some UV (string) scale where the higher derivative corrections start to dominate.

\subsection{Summary of results}
\label{sec:summaryresult}

We record here the final results of our analysis for quick reference. We find a parameterization of our entropy functional in terms of a scalar function evaluated on spatial sections of the future horizon, denoted $\sd$. To wit, 
\begin{equation}
S_{\text{total}} =  \int_{\Sigma_v}\, d^{d-2}x\, \sqrt{h} \; \Big( 1 + \sd \Big) \,.
\label{eq:parsol}
\end{equation} 
Our parameterization is such that the equilibrium contribution from the Wald analysis is subsumed into $\sd$. 
 In particular, $\sd = \sd_\text{Wald} + \sd_\text{cor}$, where $ \sd_\text{cor}$ denotes the correction terms we add which involve a series of gradient corrections suppressed by $\ell_s \omega$, as will be made more explicit below.\footnote{ While this is still a gradient expansion, it is valid as long as the low energy effective field theory makes sense. It differs from the fluid/gravity regime in that the frequencies are only constrained to be smaller than the UV scale set by $\ell_s$ and not an IR scale like temperature, which is set by local thermal equilibrium.}  We will derive the following results:
\begin{enumerate}
\item The general answer for the entropy function to all orders in $\ell_s\omega$ for Lovelock theories can be expressed as  ($\sd_\text{total} = 1 + \sd$)
\begin{equation}
\begin{split}
\sd_\text{total} =  \frac{\delta {\cal L}_\text{grav}}{\delta R^v{}_v{}^r{}_r} \bigg{|}_{R\rightarrow {\cal R}} \,
+ \sum_{n=0}^\infty \kappa_n
 \left[ \ell_s^n \partial_v^n \left( \frac{1}{2}\, \frac{\delta^2 {\cal L}_\text{grav}}{
 \delta R^A{}_{A_1}{}^{C_1}{}_v \, \delta R^{v}{}_{B_1}{}^{D_1}{}_{B} } \bigg{|}_{R\rightarrow {\cal R}} \, \Kt_{A_1}^{C_1}\Kn_{B_1}^{D_1}
\right) \right]^2 
\end{split}
 \label{eq:SfinalIntro}
 \end{equation}
 $\mathcal{L}_\text{grav}$ is the purely geometric part of the action comprising only of the metric contributions.  The replacement rule ${R\rightarrow {\cal R}} $ indicates that once we vary the Lagrangian, we replace all the curvature tensors of the spacetime with those intrinsic to the hypersurface $\Sigma_v$ (see Eq.~\eqref{defbkint} below). 

 The coefficients $\kappa_n$ are some arbitrary $\mathcal{O}(1)$ constants satisfying the following recursive inequalities\footnote{ A particular solution is provided by auxiliary parameters   $A_{-4}=\frac{1}{3},\,  A_{-3} = -\frac{1}{12}$, so that $ A_{-2} = -4,\, A_n = -1$ for $n \neq -1$ which leads to the values of the parameters $\kappa_n$ being:
 $\kappa_{-2} =-\frac{1}{2} ,\, \kappa_{-1}=-2 ,\, \kappa_{n} = -1$ for $n\geq 0$ (and $\kappa_{-3} = -1$ for consistency).} 
\begin{equation}\label{condsgb}
\begin{split}
A_{n} &=2 \,\kappa_{n}  -\frac { \kappa_{n-1}^2 }{ A_{n-2}}  \le  0\,, \qquad \text{for}\; n =-2,-1,0, \cdots. \\
& \text{initial condition}: \quad \kappa_{-2} = -\frac{1}{2} \,, \quad \kappa_{-1} = -2 \,.
\end{split}
\end{equation}
By construction, it satisfies {\bf C2}. Furthermore, for  $SO(d-1)$ spherically symmetric, time dependent geometries, it also satisfies {\bf C1}.  

\item In general Eq.~\eqref{eq:SfinalIntro} is an entropy function satisfying {\bf C1}, provided a particular total derivative term (see Eq.~\eqref{eq:Xobsgen}), which we call the \emph{obstruction term} can be bounded to be negative semidefinite. We find that the obstruction vanishes for spherically symmetric configurations but have not been able to  gain control of this term efficiently.

\item The first term in Eq.~\eqref{eq:SfinalIntro} is just the Dong-Camps (or Jacobson-Myers functional) appearing in the holographic entanglement entropy computations, see \cite{Fursaev:2013fta, Dong:2013qoa,Camps:2013zua}. This owes to our  replacement of the spacetime curvatures by the intrinsic ones on the codimension-2 hypersurface.\footnote{ As noted first by \cite{Jacobson:1993xs} this also works for the stationary black holes where the extrinsic curvatures vanish on the bifurcation surface of the horizon.} As noted earlier, \cite{Wall:2015raa} has earlier demonstrated that this functional serves to uphold the second law to linear order in amplitudes, in arbitrary higher derivative theories. Thus to leading order in the expansion our result is consistent with \cite{Wall:2015raa}.

\item More generally, the structure of Dong functional also has contributions that resemble the second term of Eq.~\eqref{eq:SfinalIntro}. They involve two variations of the Lagrangian with respect to the curvature, contracted with the extrinsic curvatures. However, despite superficial similarities, there are some subtle differences, especially in the index structure of correction term (and, in particular, the fact that we organize it in the form of a negative semidefinite quadratic form). Nevertheless, given the result of \cite{Wall:2015raa}, it is interesting to further consider whether one can find another entropy function based on the Dong functional.

\item To exemplify the answer, we can restrict to Gauss-Bonnet theory,  where we provide an explicit expression for $\sd_2$ to all orders in $\ell_s\omega$.  Explicitly, our solution reads\footnote{ For the most part we will work in a coordinate space representation where $\partial_v$ will play the role of the frequency $\omega$.}
\begin{equation} \label{scor}
\begin{split}
\sd_2&= 2\, \alpha_2 \,\ell_s^2\;\delta^{A_1 B_1}_{C_1 D_1}\; {{{\mathcal{R}^{C_1}}_{A_1}}^{D_1}}_{B_1}
+\sum_{n=0}^{\infty}\kappa_n \; \ell_s^{2n}\
 \partial_v^n \HnG{0}^A_B \;  \partial_v^n  \HnG{0}^B_A \\
\end{split}
\end{equation}
This is written in terms of data on the surface $\Sigma_v$ (see \S\ref{sec:notation} for a summary of our notation):
\begin{equation}\label{defbkint}
\begin{split}
&{\cal R}_{ABCD}=\text{intrinsic Riemann tensor on}\ \Sigma_v \
(\text{associated with}\ h_{AB})
 \,,\\
& \HnG{0}^{A}_{B} = \alpha_2 \, \ell_s^2\; \delta^{ A A_1 A_2}_{ BB_1 B_2} \;
 \Kt^{B_1}_{A_1}~\Kn^{B_2}_{A_2}, \\
&\Kt_{AB}= \frac{1}{2}\partial_v h_{AB}\vert_{r=0}, \qquad
\Kn_{AB}= \frac{1}{2}\partial_r h_{AB}\vert_{r=0}
\end{split}
\end{equation}
The first term in Eq.~\eqref{scor} agrees with the expression for Wald entropy in the  Gauss-Bonnet theory in equilibrium.
\item When the evolution breaks $SO(d-1)$ spherical symmetry, we find that  $S_{\text{total}}$  still satisfies {\bf C1} to all order in $\ell_s\omega$,
provided the following total derivative term can be bounded to be negative semidefinite:
\begin{equation}\label{expJ0int}
\begin{split}
(\partial^2_v \sd_2)_{\nabla} &= 4\, \alpha_2 \, \ell_s^2\; \nabla_A\nabla_B \left[ \Kt\Kt^{AB}- \Kt^{A}_{C}  \Kt^{BC} -\frac{h^{AB}}{2}\left( \Kt^2 - \Kt_{CD} \Kt^{CD}\right) \right] .
\end{split}
\end{equation}
Here $\nabla_A$ is the covariant derivative with respect to the spatial metric $h_{AB}$ on $\Sigma_v$ and $\Kt = \Kt^C_C$. Curiously, this obstruction term only appears to enter at  the leading order in the gradient expansion $\mathcal{O}(\omega\,\ell_s)$.\footnote{ It is also worth remarking that the term is quadratic in the amplitude away from equilibrium, and thus is invisible to the earlier linearized analysis, e.g.,  
\cite{Wall:2015raa}. We are also agnostic to the particular black hole background about which we perturb in contrast to the discussion of \cite{Sarkar:2013swa,Bhattacharjee:2015qaa}.}
\end{enumerate}
%
%

\subsection{Notation}
\label{sec:notation}
\begin{itemize}
\item Spacetime indices: Lowercase Greek $\{\alpha,\beta, \cdots, \mu,\nu, \cdots\}$.
\item Spacetime metric: $g_{\mu\nu}$ and spacetime covariant derivative $D_\alpha$.
\item Indicies on co-dimension two surfaces: Uppercase Latin $\{A,B,\cdots\}$.
\item Horizon generator: $t^\mu$, $t^\mu\,t_\mu =0$. We fix
$t^\mu = \left(\frac{\partial}{\partial v} \right)^\mu$, and abbreviate  to $\partial_v$.
\item Normalizer of horizon generator $n^\mu$, $n^\mu n_\mu  = 0$ and  $n^\mu t_\mu  = 1$. We fix $n^\mu = \left(\frac{\partial}{\partial r} \right)^\mu$, and abbreviate  to $\partial_r$.
\item Horizon tangent space generated by
$e^\mu_A \equiv \left(\frac{\partial}{\partial x^A} \right)^\mu$ and abbreviated to $\partial_A$.
\item Intrinsic metric on the horizon $h_{AB}$ and associated covariant derivative $\nabla_A$
\item Extrinsic curvatures  of the horizon:
\begin{equation}
\Kt_A^B\, e_B^\nu = e^\mu_A \, D_\mu t^\nu\,, \qquad
\Kn _A^B\, e_B^\nu = e^\mu_A \, D_\mu n^\nu  \,, \qquad
\omega^A\, e_A^\nu = n^\mu \, D_\mu n^\nu
\label{eq:KKbdef}
\end{equation}
Note that $\Kt_{AB}$ vanishes in equilibrium. For the Lovelock theories we do not find a role for the H\'{a}j\'{i}\v{c}ek one-form $\omega^A$.
\end{itemize}

\section{Set-up and strategy}
\label{sec:setup}

We begin with a general discussion of coordinate choices adapted to a black hole spacetime with a regular event horizon. Specifically, we aim to establish our choice of the metric Eq.~\eqref{metricintro} as the most general form admissible. We then explain the strategy we employ, in abstract, for the construction of an entropy function.

\subsection{Horizon adapted coordinates}
\label{sec:geom}

Our interest is in dynamical  black hole spacetimes, which admit a regular event horizon.  The horizon is a distinguished codimension-1 null hyperspace, $\mathcal{H}^+$, which is the boundary of the casual past of future null infinity. Being the boundary of  a casual set, it is  null, and is ruled by null geodesics. Our interest is when this hypersurface is dynamical, evolving non-trivially under dynamics  governed by the action Eq.~\eqref{eq:action1}.

We first choose coordinates, so that  $\mathcal{H}^+$ lies on a constant coordinate locus, w.l.o.g., say $r=0$. The null generators of the horizon will be taken to be the vector field $t^\mu = (\partial_v)^\mu$. This in turn implies that $\partial_v$ is normal to  every tangent vector field on the horizon, including $\partial_v$ itself.  We choose the coordinate $v$ such that it is the affine parameter along these null generators. Then constant $v$ slices on the horizon are a $(d-2)$ dimensional spatial manifold, denoted $\Sigma_v$. On this spacetime codimension-2 hypersurface we choose  $(d-2)$ vector fields  $e_A^\mu = (\partial_A)^\mu$ to span the tangent space. Integral curves for $\partial_A$ give us the spatial coordinates $x^A$ along $\Sigma_v$.  The vectors $\partial_A$  are orthogonal to $\partial_v$, so the metric on $\mathcal{H}^+$ takes the following degenerate form, appropriate for a null hypersurface:
\begin{equation}\label{hormet}
ds^2\vert_{\mathcal{H}^+} =  e^A_\mu \, e^B_\mu \, h_{AB}\, dx^\mu dx^\nu = h_{AB}(v,{\bf x}) ~dx^A dx^B \,,
\end{equation}

Now that we have the geometry of the future event horizon, we can construct the spacetime metric in its vicinity.  To define the coordinate $r$,  we construct a  congruence of null geodesics piercing through $\mathcal{H}^+$, at an angle fixed by the choices of the inner products between the vector field $n^\mu =(\partial_r)^\mu$ and the tangent vectors of the horizon, (i.e., $\partial_v$ and $\partial_A$). We make the  following choice:
\begin{equation}
    n^\mu t_\mu = (\partial_r,\partial_v) \Big|_{\mathcal{H}^+} =1 \,,
  \qquad
  n_\mu e_A^\mu =  (\partial_r,\partial_A)\Big|_{\mathcal{H}^+}=0
\label{eq:normalize}
\end{equation}
With this choice, the spacetime metric $g_{\mu\nu}$ in the vicinity of the horizon takes the form:
 \begin{align}\label{metch2}
 ds^2 &=    2 \,j(r,v,{\bf x}) \,dv \, dr + 2\,J_A(r,v,{\bf x}) \,dr \, dx^A -f(r,v,{\bf x}) \,dv^2 \nonumber \\
 & \qquad \qquad + 2\,k_A (r,v,{\bf x}) \, dv \, dx^A+
  h_{AB}(r,v,{\bf x}) \,dx^A dx^B\,,
 \nonumber \\
j(r,v,{\bf x})\big|_{\mathcal{H}^+}& \equiv 1\,, \qquad  J_A(r,v,{\bf x})\big|_{\mathcal{H}^+} =  f(r,v,{\bf x}) \big|_{\mathcal{H}^+}= k_A (r,v,{\bf x}) \big|_{\mathcal{H}^+}=0 \,.
 \end{align}

 By construction we have chosen $\partial_r$ to be a null vector in Eq.~\eqref{metch2}. Further imposing that null geodesics along $\partial_r$ be affinely parametrized, demands that the $r$ derivative of $j(r,v,{\bf x})$ and $J_A(r,v,{\bf x})$ vanish everywhere, implying
\begin{equation}
j(r,v,{\bf x}) =1\,, \qquad J_A(r,v,{\bf x}) =0 \,.
     \label{parraffcond}
\end{equation}
Finally, we make the choice that $v$ be the affine parameter along the null geodesics  with tangent $\partial_v$. Consequently,  we set $\partial_r f(r,v,{\bf x})\vert_{\mathcal{H}^+}=0$. Upon imposing all these choices, we arrive at the metric as described earlier in Eq.~\eqref{metricintro}, which we summarize here for convenience:
\begin{equation}
\begin{split}
    ds^2  &= 2\, dv~ dr - f(r,v, {\bf x})\,dv^2 + 2 \,k_A (r,v, {\bf x}) \,dv~ dx^A + h_{AB}(r,v, {\bf x})\,dx^A dx^B\,, \\
    & \qquad f(r=0,v, {\bf x})= k_A (r=0,v, {\bf x})= \partial_r f(r=0,v, {\bf x})=0\,.
\end{split}
\label{eq:metric}
\end{equation}
The detailed steps leading to the above constraints are given in Appendix \ref{appC}.

\subsection{Strategy of the proof}
\label{sec:strat}

As it has been already mentioned in  \S\ref{sec:state}, we would like to prove the existence of an entropy function satisfying
$$\frac{\partial S_{\text{total}}}{\partial v} \ge 0 \,,$$
for all finite $v$. While this a-priori only requires that we present a quantity which can be evaluated on different $\Sigma_v$ slices, we are going assume this is in turn defined in terms of a local density function $\Theta$.\footnote{ We might actually have desired the existence of local entropy current, a much stronger requirement, based on the fluid/gravity intuition. The dual of the entropy current would be a $(d-2)$-form which we could have integrated over arbitrary slices of the horizon, without  a-priori picking a foliation as we have done. See \S\ref{sec:discussion} for further comments.}  We write:
\begin{equation}
\label{deftheta}
\frac{\partial S_{\text{total}}}{\partial v} =\int_{\Sigma_v} d^{d-2}x \, \sqrt{h}\; \Theta\,.
\end{equation}
In  Einstein-Hilbert theory $\Theta$ is the the expansion of the null congruence along $\partial_v$. 

At this point, we simply need to show $\Theta \ge  0$. However, in analogy with the proof of the area increase theorem in Einstein-Hilbert theory, we are going to follow a different strategy. Rather than bound $\Theta$, we are going to show that within the validity of our higher derivative perturbation scheme, we have
\begin{equation}\label{restrtheta}
\partial_{v}\Theta \le 0 \,,
\end{equation}
up to the required order in derivative expansion.  Thus $\Theta$ is a monotonically decreasing function of the horizon time, $v$. Further, assuming that  in the  future the spacetime approaches an equilibrium  configuration, and hence $\Theta$  vanishes, i.e.,
\begin{equation}
\lim_{v\to \infty} \Theta(v) \; \rightarrow \;0 \,,
\end{equation}
we shall conclude that $\Theta$ is positive at every finite $v$.  Thence Eq.~\eqref{deftheta} implies that the rate of change of total entropy, $\partial_vS_{\text{total}}$, is monotone non-decreasing. We therefore can conclude that  $S_{\text{total}}$  increases with time.

Let us recall how this strategy works in Einstein-Hilbert theory Eq.~\eqref{einhilactn0} to prove the area theorem. Firstly, one constructs $S_{\text{total}} = \int_{\Sigma_v} \, d^{d-2}x\, \sqrt{h}$ and obtains therefrom
\begin{equation}
\begin{split}
\Theta_{Einstein} &= \Kt^A_A =  \frac{\partial \log\text{Area}(\Sigma_v)}{\partial v} \,, \\
\qquad \Kt_{AB} &= \frac{1}{2} \,t^\mu D_\mu h_{AB} = \frac{1}{2} \partial_v h_{AB}  \,.
\end{split}
\label{eq:defB}
\end{equation}
Note that $\Kt_{AB}$ is the Lie drag of the horizon spatial metric along the null generators and therefore it vanishes in equilibrium.\footnote{ An equilibrium black hole is a stationary solution. By virtue of stationarity, a compactly generated horizon is also a Killing horizon, generated by  a Killing vector field, $t^\mu$. Furthermore, this field has a vanishing locus on a spacetime codimension-2 surface, the bifurcation surface which lies on the horizon. This has the advantage that we can  evaluate various quantities on the bifurcation surface and then Lie drag them along $t^\mu$, which leaves them unchanged everywhere else on slices of $\mathcal{H}^+$. We should also hasten to add that stationarity alone does not imply that the horizon is Killing when the horizon is non-compact, as exemplified by  black funnel solutions (see \cite{Marolf:2013ioa} for a review and references). This will not bother us since we start from an equilibrium configuration where the horizon will be stationary and Killing, even if non-compactly generated.  }

From here it is a simple geometric computation to show that the rate of change of $\Theta$ is given by (see e.g., \cite{wald2010general})
\begin{equation}
\begin{split}
\partial_v\Theta_{Einstein} &= -\Kt_{AB} \Kt^{AB}- R_{vv}\,, \\
& = -\Kt_{\langle AB\rangle} \Kt^{\langle AB\rangle} - \frac{1}{d-2} \, \Kt^2 - R_{vv} \,,
\end{split}
\label{incrEina}
\end{equation}
where $R_{vv} =  R_{\mu\nu} (\partial_v)^\mu (\partial_v)^\nu$. All indices are raised and lowered with the metric $h_{AB}$ on $\Sigma_v$ which is defined in Eq.~\eqref{hormet}.
The reader will recognize this to be the famous Raychaudhuri equation; we have split the extrinsic curvature tensor $\Kt_{AB}$ into a symmetric-traceless shear  $\Kt_{\langle AB\rangle} $, and a scalar expansion $\Kt$, via:  
\begin{equation}
\Kt_{AB} \equiv \Kt_{\langle AB\rangle}  + \frac{1}{d-2}\, h_{AB} \, \Kt
\,. 
\label{eq:shearexp}
\end{equation}	

Up to now we have not made use of the actual dynamical equations of motion. Using Einstein's equations, we can express the result in terms of the matter energy-momentum tensor:
\begin{equation}
T_{\mu\nu} \equiv - \frac{1}{\sqrt{-g}} \frac{\delta \left(\sqrt{-g}\, \mathcal{L}_{matter}\right) }{\delta g^{\mu\nu}} \,.
\label{eq:Tdef}
\end{equation}
\begin{equation}
\partial_v\Theta_{Einstein} = -\Kt_{AB} \Kt^{AB}- T_{vv},
\label{}
\end{equation}
where we accounted for $g_{vv}\vert_{\mathcal{H}^+} =0$. We will assume that the matter is sensible, and in particular, the energy-momentum tensor satisfies the null energy condition (NEC), viz.,
\begin{equation}
 T_{\mu\nu} \xi^\mu \xi^\nu \geq 0 \,, \;\; \text{for} \;\;g_{\mu\nu} \xi^\mu\,\xi^\nu =0\,
\quad \;\; \Longrightarrow \;\;
T_{vv}   \geq 0 \,.
\label{eq:NEC}
\end{equation}
With the assumption of the null energy condition, we can immediate see that
\begin{equation}
\partial_v \Theta_{Einstein}  = -\Kt_{AB}\, \Kt^{AB} - T_{vv}
\leq 0 \,, \qquad \text{(assuming NEC)}
\label{eq:incrEin}
\end{equation}
which is the result we seek. This then implies the area theorem as detailed above.

The obvious obstacle in extending this result to higher derivative theories lies  in exchanging $R_{vv}$, via the equations of motion, with the energy momentum tensor. In general Eq.~\eqref{eq:incrEin} will have higher order terms proportional to $\alpha$, the coupling constants of the higher derivative action (and we have to also correct for the equilibrium entropy).

At this point,  the reader might be lulled into thinking that since we are working in $\omega\ell_s$ perturbation theory, so as long as the leading terms contained in the r.h.s. of Eq.~\eqref{eq:incrEin} are non-zero, the higher derivative corrections should be irrelevant. Naively, the argument goes, that lower order terms in perturbation theory should continue to dominate.  However, this logic is fallacious, for it can  happen that under the course of evolution, the r.h.s. of Eq.~\eqref{eq:incrEin} can become of $\mathcal{O}(\omega^2 \ell_s^2)$, or worse still, vanish. Recall that $\Kt_{AB}$ vanishes in equilibrium, so it is at least of $\mathcal{O}(\omega\ell_s)$, but it could indeed be smaller, depending on the particularities of the evolution.

A simple illustration of this scenario is provided by linear (in amplitude $\amp$) departures from equilibrium. Then the r.h.s. of Eq.~\eqref{eq:incrEin} is $\mathcal{O}(\amp^2)$, but higher derivative terms can, and do, contribute at  linear order $\mathcal{O}(\amp)$. In fact, demanding that such linear terms vanish, fixes some ambiguities in the Wald functional, called the JKM ambiguities after \cite{Jacobson:1993vj}, as has been described in \cite{Sarkar:2013swa,  Bhattacharjee:2015yaa, Wall:2015raa}.\footnote{  In particular, \cite{Wall:2015raa}  demonstrates how one can always ensure the absence of such linear entropy production contributions by working with the Wald-Dong functional.}

Working then to  quadratic order in $\amp$, we also have contributions from the r.h.s.  of Eq.~\eqref{eq:incrEin} and  this generically dominates over all higher derivative corrections. 
However, there are still exceptions, for  in course of  evolution, at a given point of time $\Kt_{AB}$ can become of order ${\cal O}\left(\amp\,\omega\ell_s\right)$, but its derivatives remain unsuppressed. To exemplify the situation,  consider a hypothetical case where for Wald entropy (with appropriate corrections to handle the terms linear in $\amp$), we find
$$\partial_v\Theta_{Wald} = -h^{AA'} h^{BB'}\Kt_{A'B'}\left(\Kt_{AB} +\alpha 
\,  \ell_s^2~\partial_r\partial_v \Kt_{AB}\right) - T_{vv} \,.$$
In the above equation, the ${\cal O}(\alpha \ell_s^2)$ correction term can dominate over the original $\Kt_{AB}$ term, if at a given point of time and in its neighborhood,  $\Kt_{AB} \sim {\cal O}(\alpha \ell_s^2)\, \partial_r\partial_v \Kt_{AB}$. Such a situation is not forbidden from appearing under evolution, and could clearly overwhelm the leading order Einstein-Hilbert contribution. 

Our goal is to find a way to handle such situations.  We need to correct the Wald entropy so that situations as those sketched above, do not pertain. The strategy will be to add the minimal set of corrections necessary so that the net contribution of all terms ensures non-negative definite entropy production. We will do this by suitably combining the correction terms and contributions from the Wald entropy, so that  $\partial_v\Theta $  can be expressed as a sum of perfect squares with an overall negative sign.\footnote{ This strategy, as the astute reader might appreciate, precisely follows the logic for construction of an entropy current in hydrodynamics, cf., \cite{Bhattacharyya:2008xc}.}

Before getting into the details, let us record a few salient points of our method:
 \begin{enumerate}
 \item Our construction will be perturbative in  $\omega\ell_s$, which is a natural parameter for the gradient expansion. We make no assumption about the amplitude   of the time dependent perturbation. It therefore is {\it not} an expansion around any given static or stationary background solution of the full higher derivative theory.  Consequently, there  is no way that the form of our corrections, or their coefficients, could depend on any details of a particular background solution.
 \item Note that what we really want  to prove is just $\partial_v S_{\text{total}}\geq 0$. The strategy we are employing in fact demands something \emph{ much stronger.} In a nutshell, we are demanding
 \begin{itemize}
 \item $\partial_v^2 S_{\text{total}} \le 0$ or $\partial_v S_{\text{total}}$ to be monotonically decreasing
 \item Assuming $\partial_v S_{\text{total}}$ vanishes as $v\rightarrow\infty$, this implies that $\partial_v S_{\text{total}}$ is positive for all finite $v$.
 \end{itemize}
 \item  While this proof-method is standard (owing, in part,  to the way the area theorem is proved), we must nevertheless admit that our inability to find such a function is by no means a counter-example to an entropy increase theorem. After all, it is possible for a function to be positive definite, but still not monotonically decreasing. In other words, although the monotone decrease of $\partial_v S_{\text{total}}$ is a sufficient condition for entropy increase, it is definitely not a  necessary one.  This fact is a bit frustrating since it precludes us from making strong statements about the nature of low energy effective field theory. We cannot rule out a class of higher derivative terms, and thus place constraints on the low energy limits of quantum gravity, just on the basis that our proof-method fails to construct an entropy function for them.  
 \end{enumerate}
 %

\section{Analysis for Gauss-Bonnet theory}
\label{sec:GaussBonnet}

We will begin our discussion with the Gauss-Bonnet theory,
 \begin{equation}
\begin{split}
I &=\frac{1}{4\pi} \, \int d^dx\, \sqrt{-g} \left(R + \alpha_2\, \ell_s^2\, \mathcal{L}_2 + \mathcal{L}_{matter}\right)\,,\\
\mathcal{L}_2 &= R^2 - 4 R_{\mu\nu} R^{\mu\nu} +R_{\mu\nu\alpha\beta} R^{\mu\nu\alpha\beta}=\delta^{\mu_1 \nu_1\mu_2 \nu_2}_{\rho_1 \sigma_1\rho_2 \sigma_2}~ {{{R^{\rho_1}}_{\mu_1}}^{\sigma_1}}_{\nu_1}\; {{{R^{\rho_2}}_{\mu_2}}^{\sigma_2}}_{\nu_2} \,.
\end{split}
\label{GB1}
 \end{equation}
The generalized Kronecker symbol is defined in \S\ref{sec:summaryresult} (below Eq.~\eqref{scor}), and its use will help us to generalize easily to  Lovelock gravity in \S\ref{sec:lovelock}.

As explained in \S\ref{sec:intro}, we would like to construct an expression for total entropy $S_{\text{total}}$ such that in equilibrium it reduces to Wald entropy, $S_\text{Wald}$,  and once we move away from equilibrium the inequality  Eq.~\eqref{restrtheta} is satisfied. We work in the coordinate system described in \S\ref{sec:geom}. This will suffice to illustrate the general ideas, though we could equally phrase the computation in a more natural geometric language (as will be clear from the final answer).

We will  separate the discussion into  two parts. Firstly, in \S\ref{sec:deriwald}, we shall compute the Wald entropy for Gauss-Bonnet action and we shall see that  it does not satisfy Eq.~\eqref{restrtheta} once we depart from equilibrium. From this analysis, we shall conclude that we need to correct the Wald entropy.  Then in the second part, in  \S\ref{sec:scorrGB},  we shall give our proposal for the correction, and show how, in the course of time evolution, the corrected entropy function does satisfy the inequality Eq.~\eqref{restrtheta}, provided a very particular total derivative term vanishes.
We will then explain the circumstances in which this obstruction term does not pose a problem and the lessons to be gleaned from it.  

\subsection{Part I: Time variation  of Wald entropy}
\label{sec:deriwald}

Let us first compute the time derivative of Wald entropy along spatial sections of the horizon, assuming that the black hole is dynamical  (i.e., $\partial_v$ is not a Killing vector). The expression for the equilibrium Wald entropy follows from Eq.~\eqref{GB1}, using the technique of \cite{Wald:1993nt, Iyer:1994ys}. This reads:
\begin{equation}
S_\text{Wald}= \int_{\Sigma_v} d^{d-2} x\, \sqrt{h}\; (1+\alpha_2 \,\ell_s^2\;  \sd_{2,\text{eq}}) \,.
\label{enteq}
\end{equation}
For the Gauss-Bonnet theory this evaluates to  (see e.g., \cite{Jacobson:1993xs})
\begin{equation}
\sd_{2,\text{eq}} =  2 \, \delta^{A_1 B_1}_{C_1 D_1}\; {{{R^{C_1}}_{A_1}}^{D_1}}_{B_1}
\stackrel{\text{equilibrium}}{=} \; \;  2 \,   \delta^{A_1 B_1}_{C_1 D_1}\; {{{{\cal R}^{C_1}}_{A_1}}^{D_1}}_{B_1}.
\label{defrho1}
\end{equation}
In the above equation, all the Riemann tensors are projected on the spatial sections $\Sigma_v$.  For stationary solutions, the projection will result in the  intrinsic Riemann tensor of $\Sigma_v$\footnote{  It is important to distinguish between an intrinsic quantity, denoted by an italicized symbol like $\mathcal{R}_{A_1B_1 C_1 D_1}$, and the non-intrinsic  $R_{A_1B_1 C_1 D_1}$ which depends on the full spacetime data.}, and therefore the second equality follows.

We recall that in equilibrium, $\partial_v$ is a Killing vector field, which guarantees that motion between two slices $\Sigma_v$ and $\Sigma_{v'}$ is achieved by Lie drag along a symmetry direction, which results in no changes. Let us now turn to the case where  $\partial_v$ is no longer a Killing vector. The second equality of Eq.~\eqref{defrho1} does not hold anymore and there is always an ambiguity of how to lift the equilibrium Wald entropy to the non-equilibrium situation \cite{Jacobson:1993vj}.

However,  $\sd_{2,\text{eq}}$ is only the starting point of our construction. In a time-dependent situation we anyway have to add some corrections to $\sd_{2,\text{eq}}$; these corrections will necessarily vanish in equilibrium. While the form of the correction depends on the starting point, it is clear that we can w.l.o.g.\ start with  Eq.~\eqref{defrho1}. Having obtained  a correction from this particular starting point, we can always finesse the end result, to begin from a  different initial guess for the entropy. As explained in \S\ref{sec:summaryresult}, we are therefore effectively starting with the Dong functional, which we know satisfies the second law to linear order in amplitudes, thanks to the result of \cite{Wall:2015raa}. 

With this in mind, let us assume, as our ansatz, 
that $\sd_{2,\text{eq}}$ is given by the last expression of Eq.~\eqref{defrho1} even outside equilibrium, i.e., in $\sd_{2,\text{eq}}$  all the Riemann tensors will be intrinsic:
\begin{equation}
\sd_{2,\text{eq}} \equiv   2 \,  \delta^{A_1 B_1}_{C_1 D_1}~ {{{{\cal R}^{C_1}}_{A_1}}^{D_1}}_{B_1}\,.
\label{defrho}
\end{equation}
Following  Eq.~\eqref{deftheta}, we define the equilibrium value of $\Theta$ through the temporal derivative of the Wald entropy:
\begin{equation}
\frac{\partial S_\text{Wald}}{\partial v} =\int_{\Sigma_v}\, d^{d-2} x\, \sqrt{h}~\Theta_{eq} \,.
\label{defthetaeq}
\end{equation}
The explicit expression for $\Theta_{eq}$ for the present case takes the form 
\begin{equation}
\Theta_{2,\text{eq}} = \Kt\,( 1 + \alpha_2 \,\ell_s^2\;  \sd_{2,\text{eq}}) - 2\, \alpha_2 \,\ell_s^2\; \left(\frac{\delta \sd_{2,\text{eq}}}{\delta h^{AB}}\right)\Kt^{AB} \,.
\label{eqtheta1}
\end{equation}
where $\Kt_{AB}$ is defined in Eq.~\eqref{eq:defB}, and indices are raised/lowered by $h_{AB}$, since we view all data as being defined on $\Sigma_v$.

Next we compute
\begin{equation}
\frac{\partial \sd_{2,\text{eq}} }{\partial h^{AB} }= 2\,\delta^{A_1 B_1 }_{C_1 D_1} \; \frac{\partial h^{D_1 F_1} }{\partial h^{AB} } {\mathcal{R}^{C_1}}_{A_1 F_1  B_1}  +  2\, \delta^{A_1 B_1 }_{C_1 D_1}\; h^{D_1 F_1} \; \frac{\partial}{\partial h^{AB} }\left( {\mathcal{R}^{C_1}}_{A_1 F_1 B_1}\right) \,.
\label{drdh1}
\end{equation}
The second term on the r.h.s is a total derivative. As long as the horizon is compactly generated, i.e.,
spatial sections $\Sigma_v$ are compact, we can safely discard this term. For non-compactly generated horizons (e.g., planar AdS black holes) we should impose a suitable fall-off condition  along the spatial directions. In any event, dropping this term, we find that we can write:
\begin{equation}
\Theta_{2,\text{eq}} = \Kt+2 \,\alpha_2 \, \ell_s^2\;
\Kt_A^C ~\delta^{A A_1 B_1}_{C C_1 D_1}~ {{{\mathcal{R}^{C_1}}_{A_1}}^{D_1}}_{B_1}\,,
\label{eqtheta}
\end{equation}
Taking another time derivative we finally arrive at:
\begin{equation}
\partial_{v}  \Theta_{2,\text{eq}} = \underbrace{\partial_{v}  \Kt }_{\text{Term 1}} +
\underbrace{2 \,\alpha_2 \, \ell_s^2\ \partial_{v} \Kt_A^C ~\delta^{A A_1 B_1}_{C C_1 D_1}~{{{\mathcal{R}^{C_1}}_{A_1}}^{D_1}}_{B_1}}_{\text{Term 2}} +  \underbrace{2\, \alpha_2 \, \ell_s^2\ \Kt_A^C ~\delta^{A A_1 B_1}_{C C_1 D_1}~\partial_{v} {{{\mathcal{R}^{C_1}}_{A_1}}^{D_1}}_{B_1}}_{\text{Term 3}}
\label{reldlth}
\end{equation}

We will now explicitly evaluate  each of the three terms on the  r.h.s of Eq.~\eqref{reldlth}.  The reader is directed to  Appendix \ref{appdtlcalc} for further  details of this calculation (and also for the intermediate steps leading to  Eq.~\eqref{reldlth} itself). We find
\begin{subequations}
\begin{align}
\text{Term 1} &= -\Kt^{AB}\Kt_{AB} \,-   T_{vv} -2\, \alpha_2\, \ell_s^2\;  \delta^{A_1 A_2 B_2}_{D_1 C_2 D_2} \left( \frac{\partial \Kt^{D_1}_{A_1} }{\partial v} + \Kt_{A_1C}\Kt^{D_1C} \right) {{{{R^{C_2}}_{A_2}}^{D_2}}_{B_2}}
\nonumber \\
&\qquad +\; 4  \, \alpha_2\, \ell_s^2\; \delta^{A_1 A_2 B_2}_{D_1 C_2 D_2}~  \nabla_{A_2} \Kt^{D_1}_{A_1}~\nabla^{D_2}\Kt^{C_2}_{B_2} \,,
 \label{cont1} \\
\text{Term 2} &= 2 \, \alpha_2\, \ell_s^2\; \delta^{A_1 A_2 B_2}_{D_1 C_2 D_2}~ \partial_{v} \Kt^{D_1}_{A_1} ~{{{\mathcal{R}^{C_2}}_{A_2}}^{D_2}}_{B_2} \,,
 \label{cont2} \\
\text{Term 3} &=  -4 \, \alpha_2\, \ell_s^2\; \delta^{A_1 A_2 B_2}_{D_1 C_2 D_2} \left[ \Kt^{D_1}_{A_1} \Kt^{D_2 F_2} {\mathcal{R}^{C_2}}_{A_2F_2 B_2}   +\nabla_{A_2} \Kt^{D_1}_{A_1}\nabla^{D_2}\Kt^{C_2}_{B_2} \right]  + \nabla_A \mathcal{X}^A\,.
\label{cont3}
\end{align}
\end{subequations}
where we introduce the \emph{obstruction term}:
\begin{equation}
\mathcal{X}^{A} = 4  \, \alpha_2\, \ell_s^2\;\delta^{A_1 A B_2}_{D_1 C_2 D_2}  \; \Kt^{D_1}_{A_1}\nabla^{D_2}\Kt^{C_2}_{B_2}  
\label{eq:XGBdef}
\end{equation}	
We will explain the issues with this term at the end of our analysis which should clarify its implications.  It is worth remarking that the first term has been manipulated  using the equations of motion of the Gauss-Bonnet theory. We, in particular, need to make use of the component of the equation projected onto $\mathcal{H}^+$ (the $vv$-component), see Eq.~\eqref{gbeom2}.

Adding up the three terms we obtain
\begin{align}
\partial_{v}  \Theta_{2,\text{eq}} &=  -\Kt^{AB}\Kt_{AB} - T_{vv} +\nabla_A \mathcal{X}^A
  \nonumber \\
   &  -2 \,\alpha_2\, \ell_s^2 \;\delta^{A_1 A_2 B_2}_{D_1 C_2 D_2} \bigg\{
	 \partial_{v}   \Kt^{D_1}_{A_1} ~ \big[{{{R^{C_2}}_{A_2}}^{D_2}}_{B_2} -  
	 {{{\mathcal{R}^{C_2}}_{A_2}}^{D_2}}_{B_2}\big] +
	  \Kt_{A_1C}\Kt^{D_1C}  {{{R^{C_2}}_{A_2}}^{D_2}}_{B_2}   
\nonumber \\	  
&	 \hspace{1.5in}  -2\,  \Kt^{D_1}_{A_1} \Kt^{D_2 F_2} {\mathcal{R}^{C_2}}_{A_2F_2 B_2}
	  \bigg\} \,.
\label{reldlth1}
\end{align}
The last term on the  r.h.s of Eq.~\eqref{reldlth1} can be further simplified using the following geometric identity
\begin{equation}
\delta^{A_1 A_2 B_2}_{D_1 C_2 D_2} \big[{{{R^{C_2}}_{A_2}}^{D_2}}_{B_2} -  {{{\mathcal{R}^{C_2}}_{A_2}}^{D_2}}_{B_2}\big]  = -2\, \Kn^{D_2}_{A_2}\,  \Kt^{C_2}_{B_2}  \; \delta^{A_1 A_2 B_2}_{D_1 C_2 D_2}\,.
\label{intcurid}
\end{equation}
In writing this expression we have introduced the second extrinsic curvature of the codimension-2 surface $\Sigma_v$, this time along the normal $n^\mu = (\partial_r)^\mu$ (the first, $\Kt_{AB}$, was defined earlier in Eq.~\eqref{eq:defB}) ):
\begin{equation}
\Kn_{AB} = \frac{1}{2}\, n^\mu D_\mu h_{AB} = \frac{1}{2}\, \partial_r  h_{AB}
\label{eq:defKa}
\end{equation}
We give a detailed derivation of  Eq.~\eqref{intcurid} in Appendix~\ref{appallcurv}.
There we also list the curvature tensors (evaluated on the horizon) for the metric coordinatized as  in  Eq.~\eqref{metricintro}.

Finally, when all the dust settles, we find that Eq.~\eqref{reldlth1} can be written in the following form 
\begin{align} 
    &\partial_{v}  \Theta_{2,\text{eq}} =  \underbrace{-\Kt^{AB}\Kt_{AB} - T_{vv}}_{\text{L1}}
+ \underbrace{\nabla_A \mathcal{X}^A}_{\text{O}}
     \nonumber \\
    &\underbrace{-2\, \alpha_2\, \ell_s^2\; \delta^{A_1 A_2 B_2}_{D_1 C_2 D_2}~ \Kt_{A_1C}\Kt^{D_1C}  {{{R^{C_2}}_{A_2}}^{D_2}}_{B_2} -4\, \alpha_2 \, \ell_s^2\; \delta^{A_1 A_2 B_2}_{D_1 C_2 D_2}~  \Kt^{D_1}_{A_1} \Kt^{D_2 F_2} {\mathcal{R}^{C_2}}_{A_2 F_2 B_2}}_{\text{L2}}  
    \nonumber \\
    & \underbrace{+4\, \alpha_2\, \ell_s^2\; \delta^{A_1 A_2 B_2}_{D_1 C_2 D_2}~\Kn^{D_2}_{A_2} \Kt^{C_2}_{B_2}  ~ \partial_{v}   \Kt^{D_1}_{A_1}}_{\text{L3}}
\label{reldlth1a}
\end{align}

We now want to show that the r.h.s. of Eq.~\eqref{reldlth1a} is negative semidefinite.
Let us analyze this expression term-wise to get some intuition. The first term being a perfect square is negative semidefinite owing to the explicit sign.
So is the  second term, once we assume the NEC Eq.~\eqref{eq:NEC}. This is indeed unsurprising, since these two terms are the only ones that show up in the proof of the area theorem, cf., \S\ref{sec:strat}. The remaining four terms in the second and third lines are of order ${\cal O}(\ell_s^2)$, and naively, each of them could have either sign. The obstruction term (O), we set aside for now.

Let us unpack the $\mathcal{O}(\ell_s^2)$ terms. We will first show that the terms in the second line of  Eq.~\eqref{reldlth1a} are always negligible compared to those in the first line. To appreciate this, consider grouping the terms, to write the first two lines as
\begin{equation}
\begin{split}
\text{L1+ L2}&=- \Kt_{P_1 P_2} \Kt_{Q_1 Q_2}\left[h^{P_1Q_1} h^{P_2 Q_2} + \alpha_2\, \ell_s^2\; M^{P_1Q_1P_2Q_2}\right] - T_{vv} \\
M^{P_1Q_1P_2Q_2} &=  2\,\delta^{A_1 A_2 B_2}_{D_1 C_2 D_2} \bigg( \delta^{P_1}_{A_1}h^{Q_1D_1}h^{Q_2P_2} {{{R^{C_2}}_{A_2}}^{D_2}}_{B_2}+2\,  \delta^{P_2}_{A_1}h^{P_1D_1}h^{D_2Q_1} {{{R^{C_2}}_{A_2}}^{Q_2}}_{B_2}
\bigg) \,.
\end{split}
\label{arg1}
\end{equation}

The r.h.s. of  Eq.~\eqref{arg1} can be viewed as a quadratic form built out of $h_{AB}$ and the curvature tensor ,defined on the $d^2$-dimensional space of two-tensors $\Kt_{AB}$. Let us pass to a coordinate gauge where $h_{AB} = \delta_{AB}$, so that the quadratic form in question is a combination of the identity and another symmetric matrix $M$. A further linear orthogonal basis transformation brings $M$ to its diagonal form $\text{diag}\{\lambda_M^{(1)}, \cdots, \lambda_M^{(d^2)}\}$, without affecting the identity part. In sum, by a bit of linear algebra we can diagonalize the quadratic form, whose components then behave as:
$$  1+ \alpha_2\, \ell_s^2\;  \lambda_M^k \,, \qquad k = 1, 2, \cdots \, d^2\,.$$
While generically $\lambda_M^k$ are all of $\mathcal{O}(1)$, but the pre-multiplicative factor of $\ell_s^2$ suppresses them, in our gradient expansion. This ensures that the quadratic form is close to the identity in this basis, and hence, we conclude that the second term  (L2) remains sub-dominant to the leading piece (L1).

This then leaves us with the term L3.  While it is also generally suppressed, being as it is of $\mathcal{O}(\ell_s^2)$, it differs from the terms in L1, by involving an explicit factor of $\partial_v \Kt_C^D$. This is dangerous; there are configurations where $\Kt_A^B$ gets anomalously small locally, without a compensating suppression of $\partial_v \Kt_C^D$.  It is even possible that this term ends up dominating with the wrong sign, leading to non-monotone decrease of
$S_\text{total}$. In other words, the problematic regime is one where locally
\begin{equation}
\begin{split}
&\Kt^{B_2}_{C_2}\sim \alpha_2\, \ell_s^2\;  \delta^{A_1 A_2 B_2}_{D_1 C_2 D_2}~\Kn^{D_2}_{A_2} ~ \partial_{v}  \Kt^{D_1}_{A_1} \,,\\
\text{or},~~
& \Kt_{AB}\Kt^{AB}\sim 4 \,\alpha_2\, \ell_s^2\;  \delta^{A_1 A_2 B_2}_{D_1 C_2 D_2} \nabla_{A_2} \left( \Kt^{D_1}_{A_1}\nabla^{D_2}\Kt^{C_2}_{D_2} \right) \,.
\end{split}
\label{compterm1}
\end{equation}
Should this come to pass, then L3 could potentially dominate over L1 in  Eq.~\eqref{reldlth1a} and change the overall sign of the expression. This would lead to a violation of Eq.~\eqref{restrtheta}. To ensure positivity in such situations we need to correct $S_\text{Wald}$, which we do by adding a correction term
 $S_\text{cor}$.

 The contributions to $S_\text{cor}$ will be engineered to be such that terms of the form L3, as well as, potential non-negligible contributions form $S_\text{cor}$ itself, combine with L1 of Eq.~\eqref{reldlth1a}, to produce a sum of squares with a negative semidefinite coefficient. Furthermore, it  will be required to vanish for a stationary geometry, so that each term in $S_\text{cor}$, by construction, will have at least one $\partial_v$.

A couple of comments about our strategy for determining $S_\text{cor}$ are in order:
\begin{enumerate}
 \item We make no claim regarding the uniqueness of $S_\text{cor}$. While we have found a particular choice, based on its efficacy in generating all order corrections (in $\alpha_2$), other choices are possible. Indeed, as presaged in \S\ref{sec:intro} there is no requirement for the entropy function to be unique.
 \item  Our construction has an obstruction in the form of the total derivative term in Eq.~\eqref{reldlth1a} (labeled as O). We have not been able to bound this term, so we can make a statement in the circumstances where this term vanishes. We will say more about this in due course (see also \S\ref{sec:discussion}).
 \item We will also see some curious similarities between the analysis herein and that in hydrodynamics \cite{Bhattacharyya:2013lha,Bhattacharyya:2014bha,Haehl:2015pja}. This is despite the  perturbative scheme being different, as well as, the fact that one  deals with a local entropy current instead of the total entropy in the latter context.
 \end{enumerate}

\subsection{Part II: Temporal gradient corrections to Wald entropy }
\label{sec:scorrGB}

We now turn to the determination of $S_\text{cor}$ for the Gauss-Bonnet theory. The final result has already been quoted in Eq.~\eqref{scor}, which we rewrite here for convenience:
\begin{equation}
    \begin{split}
     S_{\text{total}} &= S_\text{Wald} + S_\text{cor}\\
    S_\text{Wald}&=
        \int_{\Sigma_v} \, d^{d-2} x\, \sqrt{h} \left(1+ \alpha_2\, \ell_s^2\; \sd_{2,\text{eq}}\right)  \,, \quad
        \sd_{2,\text{eq}} \equiv 2\,  \delta^{A_1 B_1}_{C_1 D_1}~ {{{{\cal R}^{C_1}}_{A_1}}^{D_1}}_{B_1} \\
    S_\text{cor} &=
    \int_{\Sigma_v} \, d^{d-2} x\, \sqrt{h} ~\sd_{2,\text{cor}} \,, \quad
    \sd_{2,\text{cor}}=\sum_{n=0}^{\infty} \, \kappa_n \, \ell_s^{2n}\; \partial_{v}^n \HnG{0}^{A}_{B}~ \partial_{v}^n \HnG{0}^{B}_{A} \,.
    \end{split}
\label{scorall}
\end{equation}
 We have written the correction terms in a gradient expansion, where the perturbation parameter is $\ell_s \partial_v$ as explained earlier. The tensors $\HnG{0}^A_B$ are defined below. The coefficients $\kappa_n$ are taken to be $\mathcal{O}(1)$ numbers. Our task to infer whether a suitable choice of these can be made to render $\partial_v \Theta \leq 0$.

 In the process, we have introduced some notation which we will adhere to in the future to keep the expressions from getting cluttered. We define:
\begin{equation}
\begin{split}
\HnG{-1}^B_A & \equiv \ell_S^2\, \Kt^B_A\\
\HnG{0}^A_B &= \alpha_2 \, \ell_s^2 \, \delta^{ A A_1 A_2}_{ BB_1 B_2}\;  \Kt^{B_1}_{A_1}~\Kn^{B_2}_{A_2}  \,,  \\
\HnG{n}^A_B &= \partial_v^n \HnG{0}^A_B  \,. 
\end{split}
\label{eq:Hndef}
\end{equation}	
Recalling that $\Kt^A_B$ has an explicit $v$-derivative Eq.~\eqref{eq:defB}, it is $\mathcal{O}(\omega)$, as is $\Kn_{AB}$ in the gradient counting. By suitably suppling powers of $\ell_s$ we have ensured that $\HnG{n}$ has mass dimension $n$. This uniform notation is useful in the perturbation scheme, since $\ell_s^n \HnG{n}$ is a term that contributes at $n^{\rm th}$ order in our gradient perturbation theory. Indeed, note that we can write:
\begin{equation}
\sd_{2,\text{cor}} = \sum_{n=0}^\infty\, \kappa_n \, \ell_s^{2n}\, \HnG{n}^2 \,, \qquad \HnG{n}^2 \equiv \HnG{n}^A_B \, \HnG{n}^B_A \,.
\label{}
\end{equation}	

With these definitions in place, we are ready to do the computation. We  need
\begin{equation}
\Theta = \Theta_{2,\text{eq}} + \Theta_{2,\text{cor}}\,, \qquad  \Theta_{2,\text{cor}} =\frac{1}{\sqrt h} \partial_v\bigg({\sqrt{h}}~\sd_{2,\text{cor}}\bigg)
\label{thetacor}
\end{equation}
In Eq.~\eqref{eqtheta} and Eq.~\eqref{reldlth1a} we already have the contributions to $\Theta_{2,\text{eq}}$ and $\partial_v\Theta_{2,\text{eq}}$, respectively. All that remains is obtaining similar expressions for $\Theta_{2,\text{cor}}$ and its time derivative. At an abstract level this is easy, for by explicit differentiation:
\begin{equation}
\begin{split}
 \partial_v \Theta_{2,\text{cor}} =  \sd_{2,\text{cor}}\; \partial_v \Kt+ \Kt \;  \partial_v \sd_{2,\text{cor}} + \partial_v^2 \sd_{2,\text{cor}} \,.
\end{split}
\label{GBthetaf}
\end{equation}
Using our ansatz for $\sd_{2,\text{cor}}$ in Eq.~\eqref{scorall}, each term in Eq.~\eqref{GBthetaf} can be separately computed 
\begin{equation}
\begin{split}
\Kt\; \partial_{v} \sd_{2,\text{cor}} &= 
	2 \, \Kt \, \sum_{n=0}^{\infty} \kappa_n\,  \ell_s^{2n}\;   
	\HnG{n+1}^{A}_{B} \; \HnG{n}^{B}_{A} \\
\sd_{2,\text{cor}}\,  \partial_{v}\Kt &= 
	 \partial_{v}\Kt^C_C\,  \sum_{n=0}^{\infty} \kappa_n\, \ell_s^{2n}\; \HnG{n}^2 \\
\partial_{v}^2 \sd_{2,\text{cor}} &=
	 2 \, \sum_{n=0}^{\infty}\kappa_n\,  \ell_s^{2n}\; \bigg\{  \HnG{n+1}^2 + \HnG{n}^{A}_{B} \; \HnG{n+2}^{B}_{A} \bigg\} \,.
    \end{split}
 \label{relscorall}
\end{equation}

Putting together the contributions to Eq.~\eqref{relscorall} and Eq.~\eqref{reldlth1a} we finally obtain an expression for $\partial_v\Theta$:
\begin{equation}
\begin{split}
\frac{d\Theta}{d v}
& = 
-\Kt_{AB} \Kt^{AB} -T_{vv} +  \nabla_{A} \mathcal{X}^{A} 
 \\
&\quad
	  -4  \, \ell_s^{-2}\; \HnG{1}^{A_2}_{C_2}\, \HnG{-1}^{C_2}_{A_2}  + 
	4 \,\tilde{\kappa}_0 \, \ell_s^{-2}\; \HnG{0}^2 
 \\
 & \quad 
	+ \sum_{n=0}^\infty 2\, \kappa_n \, \ell_s^{2n} \bigg\{ 
	\HnG{n+1}^2 + \HnG{n}^{A}_{B} \; \HnG{n+2}^{B}_{A}
	\bigg\}
+ (\partial_v \Theta)_\text{neg}
\end{split}
 \label{dthall1}
 \end{equation}
where 
\begin{equation}
\begin{split}
 (\partial_v \Theta)_\text{neg} &  
= - 4\,  \tilde{\kappa}_0 \, \ell_s^{-2}\, \HnG{0}^2   -2 \, \alpha_2 \, \ell_s^2\;
\delta^{A_1 A_2 B_2}_{D_1 C_2 D_2}
\bigg\{ \Kt_{A_1C}\Kt^{D_1C} \,  {{{R^{C_2}}_{A_2}}^{D_2}}_{B_2} 
 \\ 
 & \hspace{1in}+2\,   \Kt^{D_1}_{A_1} \Kt^{D_2 F_2} {\mathcal{R}^{C_2}}_{A_2 F_2 B_2} - 
  2 \, \Kt^{D_1}_{A_1} \Kt^{C_2}_{A_2} \partial_{v}\Kn^{D_2}_{B_2} \bigg\} \,  \\ 
& +  \partial_{v}\Kt^C_C\,  \sum_{n=0}^{\infty} \kappa_n\, \ell_s^{2n}\; \HnG{n}^2 \\
 &  + 2 \, \Kt \, \sum_{n=0}^{\infty} \kappa_n\,  \ell_s^{2n}\;   
	\HnG{n+1}^{A}_{B} \; \HnG{n}^{B}_{A} 
 \end{split}
\label{pvTignore}
\end{equation}	

Various simple algebraic manipulations have been performed in the process of getting to Eq.~\eqref{dthall1}. Firstly, we simplified the L3 term of Eq.~\eqref{reldlth1a}  using the identity
\begin{equation}
\begin{split}
-\text{L3} &= 4\,\alpha_2 \ell_s^2 \left( \partial_{v} \Kt^{D_1}_{A_1} \right) \, \delta^{A_1 A_2 B_2}_{D_1 C_2 D_2}~\Kn^{D_2}_{B_2} \Kt^{C_2}_{A_2} 
\\
 &= 4 \, \HnG{1}^{A_2}_{C_2}\;  \Kt^{C_2}_{A_2}   -\; 4 \, \alpha_2 \,\ell_s^2\;   \delta^{A_1 A_2 B_2}_{D_1 C_2 D_2}\,  \Kt^{D_1}_{A_1} \, \Kt^{C_2}_{A_2} \,  \partial_{v}\Kn^{D_2}_{B_2}\,.
\end{split}
\label{simpli}
\end{equation}
We then supplied and removed a term proportional to $\tilde{\kappa}_0$ in the expression. One part we have left explicit in Eq.~\eqref{dthall1}, and the other part we subsumed into $ (\partial_v \Theta)_\text{neg} $. This will be helpful to write the final expression in a neat form.  

Let us understand some of the structure here, especially the rationale for introducing 
$ (\partial_v \Theta)_\text{neg} $. We claim these are terms that can be neglected in our analysis as they will never dominate in the $(\omega\ell_s )$ perturbation theory over terms that we explicitly retain in Eq.~\eqref{dthall1}. The rationale for dropping these is as follows:
\begin{enumerate}
\item   All terms in the first two lines of Eq.~\eqref{pvTignore} have at least two powers of $\Kt_{AB}$ multiplying other tensors. As such, they are negligible compared to the leading $\Kt_{AB} \Kt^{AB}$ for reasons elaborated in \S\ref{sec:deriwald}, and so can safely be dropped.
\item The terms in the third line proportional to $\partial_v \Kt $ are controlled as follows. Let us contrast  them against the term  proportional to $\HnG{n+1}^2$ in Eq.~\eqref{dthall1}. Juxtaposing these two terms together we would have a contribution 
\begin{equation}
 \sum_{n=1}^\infty\,  \, \ell_s^{2n-2}
\left(2 \, \kappa_{n-1} + \kappa_n\, \ell_s^2 \,\partial_v\Kt \right) \HnG{n}^2 + \kappa_0\,\partial_v\Kt \,  \HnG{0}^2 \,. 
\end{equation}	
In the infinite sum, each term is dominated by the $\kappa_{n-1}$ term which is the one we have retained. The isolated  piece involving $ \kappa_0\,\partial_v\Kt \,  \HnG{0}^2$ has again two powers of $\Kt_A^B$ in the tensor $\HnG{0}$, and therefore is negligible in our expansion scheme.
\item The terms in the last line which contain $\Kt\; \HnG{n+1}^A_B \,\HnG{n}_A^B$   need more care, but they too can in the end be neglected. To appreciate this let us contrast them against the first term, $-\Kt_{AB} \, \Kt^{AB}$. Isolating just these two terms we have the sub-expression $T_1$  of Eq.~\eqref{dthall1}  
\begin{equation}
T_1 \equiv- \Kt^{B}_{A} \left[ \Kt^A_B - \delta^A_B \, \sum_{n=0}^{\infty} \,\kappa_n\, \ell_s^{2n} \, \HnG{n+1}^C_D \, \HnG{n}^D_C \right]
\end{equation}	
It is then is clear that a  term in the infinite sum will dominate only when the leading $\Kt_A^B$ becomes of order one of the term in the parenthesis. At this point we can combine them and learn that 
\begin{equation}
\begin{split}
&\Kt^A_B \sim \delta^A_B\, \kappa_n \, \ell_s^{2n}\, \HnG{n+1}^C_D \, \HnG{n}^D_C \\
& \Longrightarrow
\;  T_1 \sim \kappa_n^2 \, \ell_s^{4n} \, \left(\HnG{n+1}^A_B \, \HnG{n}^A_B\right)^2
\end{split}
\end{equation}	
However, now this term is subdominant to the $\HnG{n}^2$ term which we have retained, and thus may be safely neglected. 
\end{enumerate}

The remaining terms which we have left explicit in Eq.~\eqref{dthall1} are all the ones which we should examine closely. While  some are naively suppressed in powers of 
$\omega\ell_s$, they have the potential to reverse the sign of the entropy production by dominating over the leading order terms. Let us collect these terms into a useful form, which will enable us to bring them to a sum of perfect squares.  Let us use Eq.~\eqref{eq:Hndef}
to write the leading $\Kt_{AB}\, \Kt^{AB}$ contribution in terms of $\HnG{-1}$.
Dropping $ (\partial_v \Theta)_\text{neg} $ we have the temporal variation of entropy production being given now by 
\begin{equation}
\begin{split}
&\frac{d\Theta}{d v} =
-T_{vv}  +  \nabla_{A} \mathcal{X}^A -  \ell_s^{-4}\, \HnG{-1}^{A}_{B}\HnG{-1}^{B}_{A} - 
4 \, \ell_s^{-2}
\, \HnG{1}_{B}^{A} \HnG{-1}^{B}_{A}\\
& + 4\,\tilde{\kappa}_0\, \ell_s^{-2}\, \HnG{0}^2 +2 \sum_{n=1}^{\infty} \kappa_{n-1} \, \ell_s^{2n-2}
\HnG{n}^2 +  2 \sum_{n=0}^{\infty}\kappa_n\, \ell_s^{2n}  \HnG{n}^{A}_{B} \HnG{n+2}^{B}_{A} 
\end{split}
\label{dthall3}
\end{equation}
Furthermore, introduce  
\begin{equation}
 \label{defkaps}
\kappa_{-2} = -\,\frac{1}{2} \,,  \qquad  \tilde{\kappa}_0 = -1\,, \qquad \kappa_{-1} = -2\,.
\end{equation}
After a bit more manipulation it is possible to cast the final expression for $\partial_{v}\Theta$ as
\begin{equation}
\begin{split}
\frac{d\Theta}{d v}
&= -T_{vv}  + \mathcal{J}+  \nabla_{A} \mathcal{X}^A  \,
\\
\mathcal{J} &= 2 \sum_{n=-1}^{\infty}   \ell_s^{2n-2}  \left[
\kappa_{n-1} \, \HnG{n}^2  +   \kappa_{n}   \, \ell_s^2\, \HnG{n}^{B}_{A}  \HnG{n+2}^{A}_{B} \right]\,.
\end{split}
 \label{dthallf}
\end{equation}
We have now successfully isolated the contributions of the $\HnG{n}$ into $\mathcal{J}$ defined above. Importantly, in obtaining  Eq.~\eqref{dthallf}, we have succeeded in getting a quadratic polynomial in the tensor$\HnG{n}^B_A$. The quadratic form is given by a matrix, $\mathcal{M}$, which matrix is band-diagonal in its coefficients,  since we have links between $\HnG{n}$ and $\HnG{n+2}$ at most (and $n \geq -1$). We have
\begin{itemize}
\item the diagonal entries of $\mathcal{M}$ given by the first term on the r.h.s of $\mathcal{J}$ in Eq.~\eqref{dthallf}, viz.,  $2\, \kappa_{n-1} \, \ell_s^{2n-2}$,
\item the off-diagonal entries of $\mathcal{M}$ are in the second term on the r.h.s of $\mathcal{J}$ in Eq.~\eqref{dthallf}, viz.,  $2\kappa_{n} \, \ell_s^{2n}$, for $n = -1,0,1,2,\cdots$.
\end{itemize}
The matrix $\mathcal{M}$ can thus be explicitly constructed  
{\footnotesize
\begin{equation} \label{defmatrix}
 \mathcal{M} = \left( \begin{array}{ccccccccc}
2 \kappa_{-2} \,\ell_s^{-4} & 0 & \kappa_{-1} \,\ell_s^{-2} & 0  & 0 & 0 & 0 & \cdots \\
0  & 2 \kappa_{-1} \ell_s^{-2} & 0 & \kappa_{0}  & 0 & 0 & 0 & \cdots\\
\kappa_{-1}   \ell_s^{-2} & 0 & 2 \kappa_{0}  & 0 & \kappa_{1}  \ell_s^{2} & 0 & 0 & \cdots \\
0 & \kappa_{0}   & 0 & 2 \kappa_{1} \ell_s^{2} & 0 & \kappa_{2}  \ell_s^{4} & 0 & \cdots  \\
0 & 0 & \kappa_{1} \ell_s^{2} & 0 & 2 \kappa_{2} \ell_s^{4} & 0 & \kappa_{3}  \ell_s^{6} & \cdots  \\
0 & 0 & 0 & \kappa_{2}  \ell_s^{4} & 0 & 2 \kappa_{3} \ell_s^{6} & 0 & \cdots \\
0 & 0 & 0 & 0 & \kappa_{3}  \ell_s^{6} & 0 & \ddots & \cdots  \\
\vdots & \vdots & \vdots & \vdots & \vdots & \vdots & \vdots & \ddots
\end{array} \right) .
\end{equation}
}

As long as we can exhibit a choice of $\kappa_n$ such that the matrix $\mathcal{M}$ is a negative definite quadratic form, and thus $\mathcal{J} \leq 0$, we can declare victory. This can be done by suitably grouping terms as follows:
\begin{equation}
\begin{split}
\mathcal{J} &= \sum_{n=-2}^\infty \, \ell_s^{2n}\,  A_{n}  \left[\HnG{n+1} +
\frac {\kappa_{n+1} }{ A_{n}} \, \,\ell_s^2\, \HnG{n+3} \right]^2  
\end{split}
\label{dthallf1}
\end{equation}
where $A_{n}$ is defined through a pair of continued fractions\footnote{We would like to thank Dileep Jatkar for useful discussions about this point.}
\begin{equation}
\begin{split}
 A_{2p} &= 2\, \kappa_{2p} - { \kappa_{2p-1}^2 \over 2 \kappa_{2p-2} - { \kappa_{2p-3}^2 \over 2 \kappa_{2p-4} -{ \#\#\#\#  \over   \#\#\#\# - {\kappa_{1}^2 \over 2 \kappa_{0} -{\kappa_{-1}^2 \over 2 \kappa_{-2}} }}}}\,,\\
A_{2p-1} &= 2 \kappa_{2p-1} - { \kappa_{2p-2}^2 \over 2 \kappa_{2p-3} - { \kappa_{2p-4}^2 \over 2 \kappa_{2p-5} -{ \#\#\#\# \over   \#\#\#\# - {\kappa_{2}^2 \over 2 \kappa_{1} -{\kappa_{0}^2 \over 2 \kappa_{-1}} }}}}\,.
\end{split}
\end{equation}
It is also useful to note that they satisfy recursion relations
\begin{equation}
\begin{split}
A_{n} &=2 \kappa_{n}  - { \kappa_{n-1}^2 \over A_{n-2}} \,, \qquad \text{for}\; n =-2,-1,0, \cdots. 
\end{split}
\end{equation}

Therefore, the constraints we need are obtained demanding that
\begin{equation}
\begin{split}
&A_{n} \le 0 \,, \qquad \text{for}\; n  \geq -2\,.
\end{split}
\end{equation}
These will suffice to ensure that $ \mathcal{J}\le 0$. Furthermore,  once we assume the NEC to guarantee $T_{vv} \ge 0$, we see that $\partial_v \Theta $ in Eq.~\eqref{dthallf} is negative semidefinite, modulo the total derivative term $\nabla_A \mathcal{X}^A$. As long as we can control this term, we have achieved our stated goal of constructing a suitable entropy function for Gauss-Bonnet theory.

The total derivative term, which is our obstruction from giving a clean proof, can be re-expressed as follows: 
\begin{equation}
\begin{split}
\nabla_A \mathcal{X}^A &= \nabla_A \nabla_B \mathcal{Y}^{AB} \\
\mathcal{Y}^{AB} &= 2\, \alpha_2\, \ell_s^2  \left[ 2\Kt\Kt^{AB}- 2\Kt^{A}_{C}  \Kt^{BC} -h^{AB} \left((\Kt^{C}_{C})^2 - \Kt_{CD} \Kt^{CD}\right) \right]
\end{split}
\label{expJ0}
\end{equation}
As long as we can control this total derivative term, we have managed to construct an 
entropy function for Gauss-Bonnet gravity.

We require this  obstruction term, $\nabla_A  \mathcal{X}^A $, to be negative semidefinite. An easy way to achieve this is to in fact make it vanish. This happens, for instance, if we consider $SO(d-1)$ spherically symmetric solutions, where time evolution preserves the spherical symmetry. Owing then to the spatial symmetry, the geometric structures on $\Sigma_v$ are constrained. In particular, $\mathcal{Y}_{AB} \propto h_{AB}$ and thus the gradient term vanishes. 

More generally, for compact horizon topology, with shear-free null generators, $\Kt_{\langle AB\rangle} =0$, we can infer that  vanishing of $\nabla_A \mathcal{X}^A$ in Eq.~\eqref{expJ0}, is equivalent to demanding that $\Kt^2$ be a harmonic function on $\Sigma_v$. Since all harmonic functions on compact spaces are constants, it then follows that as long as the horizon is generated by shear-free generators $t^\mu$ with constant expansion, we would have an entropy function, since $\nabla_A \mathcal{X}^A =0$. 

We find it curious that the term of interest $\mathcal{X}^A$ appears exactly only at  the linear order in the gradient expansion. It would be interesting to characterize the solution space to $\nabla_A \mathcal{X}^A \leq 0$ more generally, a task we leave for future investigation.

\section{Higher order Lovelock terms}
\label{sec:lovelock}

Having explained how to construct an entropy function for Gauss-Bonnet gravity in some detail, we now turn to higher order Lovelock terms. The action for these theories is given by
\begin{equation} \label{eq:LoveAction}
\begin{split}
 I_m &\equiv \frac{1}{4\pi} \, \int \sqrt{-g} \; \left[ R + \alpha_m \, \ell_s^{2m-2}\,  
 \mathcal{L}_m + \mathcal{L}_{matter} \right] \\
\mathcal{L}_m &=  \delta^{\mu_1\nu_1\cdots \mu_m \nu_m}_{\rho_1\sigma_1\cdots \rho_m \sigma_m} R^{\rho_1}{}_{\mu_1}{}^{\sigma_1}{}_{\nu_1} \cdots R^{\rho_m}{}_{\mu_m}{}^{\sigma_m}{}_{\nu_m} \,.
\end{split}
\end{equation}
where $\delta^{\mu_1\nu_1\cdots \mu_m \nu_m}_{\rho_1\sigma_1\cdots \rho_m \sigma_m}$, the generalized Kronecker symbol, is defined below Eq.~\eqref{scor}.  While \S\ref{sec:GaussBonnet} focuses on $m=2$, we now explain how to extend these considerations for  $m>2$.  We will first work out the story for a single Lovelock term and then consider the general theory Eq.~\eqref{eq:action2}.

Modulo some additional tensor structures, this will work in a very similar vein, so our discussion will be brief.

\subsection{Part I: Time variation of Wald entropy}

The Wald entropy function for Lovelock theories of gravity evaluates to
\begin{equation}
\begin{split}
S_\text{Wald}^{(m)} &\equiv \int_{\Sigma_v} d^{d-2}x \,\sqrt{h} \; \left( 1 +  \alpha_m \, \ell_s^{2m-2} \, \sd_{m,\text{eq}} \right) \\
\sd_{m,\text{eq}} & =   m \, \delta^{A_1B_1\cdots A_{m-1} B_{m-1}}_{C_1D_1 \cdots C_{m-1} D_{m-1}} \,  {\cal R}^{C_1}{}_{A_1}{}^{D_1}{}_{B_1} \cdots {\cal R}^{C_{m-1}}{}_{A_{m-1}}{}^{D_{m-1}}{}_{B_{m-1}} 
\end{split}
\end{equation}	
Defining $\Theta_{m,\text{eq}}$ through the derivative of $S_\text{Wald}^{(m)}$ as in Eq.~\eqref{defthetaeq}, we find after a simple series of manipulations
\begin{equation}
\begin{split}
\Theta_{m,\text{eq}} &= \Kt( 1 + \alpha_m\, \ell_s^{2m-2}\,  \sd_{m,\text{eq}}) - 
2\, \alpha_m\, \ell_s^{2m-2}  \left(\frac{\delta \sd_{m,\text{eq}}}{\delta h^{AB}}\right)\Kt^{AB} \\
&= \Kt+ m \, \alpha_m\, \ell_s^{2m-2}  \, \Kt^C_A\;  \delta^{AA_1B_1\cdots A_{m-1} B_{m-1}}_{CC_1D_1 \cdots C_{m-1} D_{m-1}} \, \left( {\cal R}^{C_1}{}_{A_1}{}^{D_1}{}_{B_1} \cdots {\cal R}^{C_{m-1}}{}_{A_{m-1}}{}^{D_{m-1}}{}_{B_{m-1}} \right)  \,.
\end{split}
\label{eq:ThetaLove}
\end{equation}	
As in \S\ref{sec:deriwald} we will take this result which is written in terms of the intrinsic data on $\Sigma_v$ as our seed ansatz for the entropy function. This functional will satisfy the second law to linear order in amplitudes \cite{Wall:2015raa}. We  now wish to find corrections to it, so as to have a second  law for arbitrary amplitude departure away from equilibrium within the remit of effective field theory.

In the following we will make use of some abbreviations to declutter notation. In particular, let us introduce the shortcuts\footnote{ As one might expect since the Lovelock terms comprise of Euler forms in even dimensions this notation should be  naturally suggestive of the differential form structure in terms of the curvature 2-form. }
\begin{equation}
\label{eq:abbreviations}
\begin{split}
\left( {\cal R}^{m-1} \right)^A_C & \equiv \delta^{AA_1B_1\cdots A_{m-1} B_{m-1}}_{CC_1D_1 \cdots C_{m-1} D_{m-1}} \,  \left( {\cal R}^{C_1}{}_{A_1}{}^{D_1}{}_{B_1} \cdots {\cal R}^{C_{m-1}}{}_{A_{m-1}}{}^{D_{m-1}}{}_{B_{m-1}} \right) \,, \\
\left( {\cal R}^{m-2} \right)^{AA_1B_1}_{CC_1D_1} & \equiv \delta^{AA_1B_1\cdots A_{m-1} B_{m-1}}_{CC_1D_1 \cdots C_{m-1} D_{m-1}} \,  \left( {\cal R}^{C_2}{}_{A_2}{}^{D_2}{}_{B_2} \cdots {\cal R}^{C_{m-1}}{}_{A_{m-1}}{}^{D_{m-1}}{}_{B_{m-1}} \right) \,,\\
\left( {\cal R}^{m-3} \right)^{AA_1B_1A_2B_2}_{CC_1D_1C_2D_2} & \equiv \delta^{AA_1B_1\cdots A_{m-1} B_{m-1}}_{CC_1D_1 \cdots C_{m-1} D_{m-1}} \,  \left( {\cal R}^{C_3}{}_{A_3}{}^{D_3}{}_{B_3} \cdots {\cal R}^{C_{m-1}}{}_{A_{m-1}}{}^{D_{m-1}}{}_{B_{m-1}} \right)\,.
\end{split}
\end{equation}
Using this shorthand notation, we can take one more derivative of Eq.~\eqref{eq:ThetaLove}:
\begin{equation}
\begin{split}
\partial_v \Theta_{m,\text{eq}}
&= \underbrace{\partial_v \Kt }_{\text{Term 1}} + \underbrace{m \, \alpha_m \, \ell_s^{2m-2}  \, \partial_v \Kt^C_A\;  \left( {\cal R}^{m-1} \right)^{A}_{C}}_{\text{Term 2}}  \\
&\quad + \underbrace{m(m-1) \, \alpha_m \, \ell_s^{2m-2}  \,  \Kt^C_A\;   \left( \partial_v {\cal R}^{C_1}{}_{A_1}{}^{D_1}{}_{B_1} \right) \left( {\cal R}^{m-2} \right)^{AA_1B_1}_{CC_1D_1}}_{\text{Term 3}} \,.
\end{split}
\label{eq:ThetaCalc}
\end{equation}	

The there different terms can again be simplified in pretty much the same fashion as before. We have:
\begin{subequations}
\begin{align}
\text{Term 1} & =  - \Kt_{AC} \, \Kt^{AC}- T_{vv} + \alpha_m\, \ell_s ^{2m-2} \, \mathcal{E}^{(m)}_{vv} \\
\text{Term 2} &=
	 m \,  \alpha_m\, \ell_s^{2m-2}  \; \partial_v \Kt^C_A\; \left( {\cal R}^{m-1} \right)^{A}_{C}  \,, \\
\text{Term 3} &= 
	 -2m(m-1) \,  \alpha_m\, \ell_s^{2m-2}  \left[   
	  \left( \nabla_{A_1} \Kt_A^C \right) \left( \nabla^{D_1} \Kt_{B_1}^{C_1} \right) 
	  + \Kt^C_A \, \Kt_D^{D_1}  {\cal R}^{C_1}{}_{A_1}{}^{D}{}_{B_1} \right]
\nonumber \\
& \hspace{2in} \times  
	\left( {\cal R}^{m-2} \right)^{AA_1B_1}_{CC_1D_1}   
	 +  \nabla_{A} \widetilde{\mathcal{X}}_{(m)}^{A}  \,,
\label{eq:Lt123}
\end{align}	
\end{subequations}

We have made several simplifications in writing the final form of these terms. For one, 
in the first term we have used the temporal components of the Lovelock equations of motion (see Appendix \ref{app:loveDvv} for details). This reads:\footnote{  By writing the equations of motion in this form, we are assuming $m\geq 3$. If $m=2$, we can revert back to the Gauss-Bonnet case, which only differs in that the product of curvatures in the first line of Eq.~\eqref{eq:love1} is not present.}
\begin{equation}
\begin{split}
\alpha_m\, \ell_s^2  \mathcal{E}_{vv}^{(m)} &= -m \,\alpha_m\, \ell_s^{2m-2}\,   \left( {R}^{m-2} \right)^{AA_1B_1}_{CC_1D_1}  \\
  &\qquad \times \left[ \left(\partial_v \Kt_A^C + \Kt_A^{E} \, \Kt^C_E \right) {R}^{C_{1}}{}_{A_{1}}{}^{D_{1}}{}_{\Kt_{1}}- 2\,(m-1)\, \left( \nabla_{A_1} \Kt^C_A \right) \left( \nabla^{D_1} \Kt_{B_1}^{C_1} \right) \right] \,,
 \end{split}
\label{eq:love1}
\end{equation}	
The  tensors like $\left( {R}^{m-2} \right)^{AA_1B_1}_{CC_1D_1} $ are defined as in Eq.~\eqref{eq:abbreviations}, but with the projected curvature tensor, $R_{ABCD}$, instead of the intrinsic one, ${\cal R}_{ABCD}$.

For another, to evaluate the third term, we employ a strategy analogous to the discussion around Eq.~\eqref{cont3}. This leads to the total derivative piece, where the vector field $\widetilde{\mathcal{X}}_{(m)}^A$ now has the expression: 
\begin{equation}
 \widetilde{\mathcal{X}}^{A_1}_{(m)} \equiv 2\,m \, (m-1)\, \alpha_m \, \ell_s^{2m-2} \;
\Kt_A^C \, \left( \nabla^{D_1} \Kt_{B_1}^{C_1} \right)  \left( {\cal R}^{m-2} \right)^{AA_1B_1}_{CC_1D_1}   \,.
\end{equation}
Again, this total derivative term will be an obstruction to our proof aiming at giving a clean result for an entropy function.

Adding up the contributions from the three terms given above, we find in total
 \begin{align}
 &\partial_v \Theta_{m,\text{eq}} = - T_{vv} - \Kt_{AC} \, \Kt^{AC}   + 
 \nabla_A  \widetilde{\mathcal{X}}^A_{(m)} 
 - m \, \alpha_m \, \ell_s^{2m-2}\,  \bigg\{
  \partial_v \Kt_A^C \,
 \left[ \left( {R}^{m-1} \right)^{A}_{C}  - \left( {\cal R}^{m-1} \right)^{A}_{C}  \right]  
 \nonumber \\
  &\qquad\qquad\;\;- 2(m-1)  \left( \nabla_{A_1} \Kt^C_A \right) \left( \nabla^{D_1} \Kt_{B_1}^{C_1} \right)
  \left[ \left( {R}^{m-2} \right)^{AA_1B_1}_{CC_1D_1}  - \left( {\cal R}^{m-2} \right)^{AA_1B_1}_{CC_1D_1}  \right]
  \nonumber \\
 &\qquad\qquad\;\; + \Kt_A^E \, \Kt_F^G \Big[ \delta_E^F \delta_G^C \, {R}^{C_{1}}{}_{A_{1}}{}^{D_{1}}{}_{B_{1}} \left( {R}^{m-2} \right)^{AA_1B_1}_{CC_1D_1}   
 \nonumber \\
 &\qquad\qquad\qquad\qquad\qquad+2(m-1) \, \delta_E^C \delta^{D_1}_G \; {\cal R}^{C_{1}}{}_{A_{1}}{}^{F}{}_{B_{1}}\left( {\cal R}^{m-2} \right)^{AA_1B_1}_{CC_1D_1}   \Big]
\bigg\}  \,.
 \label{eq:dvthetalove1}
\end{align}

 We can further simplify this expression, we first note that the last two terms contain explicit 
 factors of two extrinsic curvature tensors, and therefore remain suppressed relative to the leading 
 $\Kt_{AC}\Kt^{AC}$ term (see the discussion around Eq.~\eqref{arg1}). This leaves us with simplifying the two terms in the first two lines which involve differences of powers of the projected and intrinsic  curvature tensors. We expand them using standard geometric identities in terms of extrinsic curvatures and then argue that all but the leading term are irrelevant. Explicitly,
\begin{align}
&\left( {R}^{m-1} \right)^{A}_{C} - \left( {\cal R}^{m-1} \right)^{A}_{C}  \nonumber \\
&\quad = \delta^{AA_1B_1\cdots A_{m-1} B_{m-1}}_{CC_1D_1 \cdots C_{m-1} D_{m-1}}  \Big[ R^{C_{1}}{}_{A_{1}}{}^{D_{1}}{}_{B_{1}} \cdots R^{C_{m-1}}{}_{A_{m-1}}{}^{D_{m-1}}{}_{B_{m-1}}   \nonumber \\
&\qquad\qquad\qquad\qquad\qquad\quad - {\cal R}^{C_{1}}{}_{A_{1}}{}^{D_{1}}{}_{B_{1}} \cdots {\cal R}^{C_{m-1}}{}_{A_{m-1}}{}^{D_{m-1}}{}_{B_{m-1}}  \Big]  \nonumber \\
&\quad= -2(m-1) \, \Kn_{A_1}^{D_1} \, \Kt_{B_1}^{C_1} \; \delta^{AA_1B_1\cdots A_{m-1} B_{m-1}}_{CC_1D_1 \cdots C_{m-1} D_{m-1}} \,  {\cal R}^{C_{2}}{}_{A_{2}}{}^{D_{2}}{}_{B_{2}} \cdots {\cal R}^{C_{m-1}}{}_{A_{m-1}}{}^{D_{m-1}}{}_{B_{m-1}}   \nonumber \\
&\qquad\, + \mathcal{O}(\Kn_{AB}\Kt_{CD})^2  \nonumber \\
&\quad = -2(m-1) \, \Kn_{A_1}^{D_1} \, \Kt_{B_1}^{C_1} \left( {\cal R}^{m-2} \right)^{AA_1B_1}_{CC_1D_1}
+ \mathcal{O}(\Kn_{AB}\Kt_{CD})^2 \,.
\end{align}
Using this formula (and a similar one for the term involving $m-2$ powers of curvature)  we can bring Eq.~\eqref{eq:dvthetalove1} to the form:
 \begin{align}
\partial_v \Theta_{m,\text{eq}} &=
	 - T_{vv} - \Kt_{AC} \, \Kt^{AC} + \nabla_A  \widetilde{\mathcal{X}}^A_{(m)}   \nonumber \\
  &
  	 +2 m(m-1) \, \alpha_m\,\ell_s^{2m-2} \, \left( {\cal R}^{m-3} \right)^{AA_1B_1A_2B_2}_{CC_1D_1C_2D_2}   \nonumber \\
  &
  	\qquad \times \Big\{
  	\partial_v \Kt_A^C \;
  	 \Kn_{A_1}^{D_1} \, \Kt_{B_1}^{C_1} \; {\cal R}^{C_{2}}{}_{A_{2}}{}^{D_{2}}{}_{B_{2}} 
  	  -2(m-2) \left( \nabla_{A_1} \Kt^C_A \right) \left( \nabla^{D_1} \Kt_{B_1}^{C_1} \right)
   	\Kn_{A_2}^{D_2} \, \Kt_{B_2}^{C_2} \Big\}  \nonumber \\
   &	+  \mathcal{O}\left(\alpha_m\, \ell_s^{2m-2} \Kt_{AC}^2\right) .
 \label{eq:dvthetalove2}   
\end{align}
Here, we discarded higher order terms which are at least of the order indicated as they are always suppressed compared to the bare $-\Kt_{AC}\Kt^{AC}$ contribution, which is negative semidefinite.

Finally, we can rewrite the second term in the curly bracket in Eq.~\eqref{eq:dvthetalove2} as a total derivative by observing that
\begin{equation}
\begin{split}
& \left( {\cal R}^{m-3} \right)^{AA_1B_1A_2B_2}_{CC_1D_1C_2D_2} \Big\{
   \left( \nabla_{A_1} \Kt^C_A \right) \left( \nabla^{D_1} \Kt_{B_1}^{C_1} \right)
   \Kn_{A_2}^{D_2} \, \Kt_{B_2}^{C_2} \Big\} \\
&\quad =  \frac{1}{2}\, \nabla_{A_1} \Big\{ \left( {\cal R}^{m-3} \right)^{AA_1B_1A_2B_2}_{CC_1D_1C_2D_2}  \Kt_A^C \left( \nabla^{D_1} \Kt_{B_1}^{C_1} \right) \!
   \Kn_{A_2}^{D_2} \Kt_{B_2}^{C_2}  \Big\} + \mathcal{O}(\Kt_{AC}^2) \,,
\end{split}
\label{eq:LRKKb}
\end{equation}

This means we can ignore it in the analysis at the expense of amending our obstruction from 
$\widetilde{\mathcal{X}}^A_{(m)} \mapsto \mathcal{X}^A_{(m)}$. All told, we find that 
 Eq.~\eqref{eq:dvthetalove2}  can be simplified to
 \begin{equation}
\begin{split}
 \partial_v \Theta_{m,\text{eq}} &= - T_{vv} - \Kt_{AC} \, \Kt^{AC}  + \nabla_A \mathcal{X}^A_{(m)} \\
  &\quad +2 m(m-1) \, \alpha_m\,\ell_s^{2m-2} \,\left( {\cal R}^{m-2} \right)^{AA_1B_1}_{CC_1D_1}\, \Big\{
  \partial_v \Kt_A^C \;
 \Kn_{A_1}^{D_1} \, \Kt_{B_1}^{C_1} \Big\}  
 \end{split}
 \label{eq:dvthetalove3}
 \end{equation}
where the total derivative term now reads
 \begin{equation}
 \begin{split}
\mathcal{X}_{(m)}^{A_1}& \equiv
2\,m \, (m-1)\, \alpha_m \, \ell_s^{2m-2} \; \\
& \times \left( {\cal R}^{m-3} \right)^{AA_1B_1A_2B_2}_{CC_1D_1C_2D_2}  \, \Kt_A^C \, \left( \nabla^{D_1} \Kt_{B_1}^{C_1} \right)  \left( {\cal R}^{C_2}{}_{A_2}{}^{D_2}{}_{B_2} - (m-2) \, \Kn_{A_2}^{D_2} \, \Kt_{B_2}^{C_2} \right)  \,.
  \end{split}
  \label{eq:LXterm}
\end{equation}
In this boundary term, the piece proportional to $(m-2)$ is new compared to the Gauss-Bonnet analysis and has resulted from Eq.~\eqref{eq:LRKKb}.

\subsection{Part II: Temporal gradient corrections to Wald entropy}

In writing the final expression for  $\partial_v \Theta_{m,\text{eq}}$ in Eq.~\eqref{eq:dvthetalove3}, we have already discarded terms which are subleading relative to the leading contribution $-\Kt_{AC}\Kt^{AC}$ (which is manifestly negative semidefinite). However, as discussed at length earlier, this term can be overwhelmed by the contribution from the second line of Eq.~\eqref{eq:dvthetalove3}. Therefore as in \S\ref{sec:scorrGB} we are going to add corrections to the Wald entropy to construct an entropy function. 

Our candidate entropy function is given as:
\begin{equation}\label{eq:LovelockEntropy}
\begin{split}
 S_{\text{total}}^{(m)} &\equiv S_\text{Wald}^{(m)} + S_\text{cor}^{(m)} \equiv 
 \int_{\Sigma_v} d^{d-2}x \, 
 \sqrt{h} \; \bigg[ \left( 1 +  \alpha_m\, \ell_s^{2m-2} \; \sd_{m,\text{eq}}  \right)+ 
  \sd_{m,\text{cor}}  \bigg]\\
 \sd_{m,\text{eq}} &= m \;  \delta^{A_1B_1\cdots A_{m-1} B_{m-1}}_{C_1D_1 \cdots C_{m-1} D_{m-1}} \,  {\cal R}^{C_1}{}_{A_1}{}^{D_1}{}_{B_1} \cdots {\cal R}^{C_{m-1}}{}_{A_{m-1}}{}^{D_{m-1}}{}_{B_{m-1}} \,,\\
\sd_{m,\text{cor}} &= \sum_{n=0}^\infty \kappa_n \, \ell_s^{2n}  \HnL{n}^A_B \HnL{n}^B_A
 \end{split}
\end{equation}
where once again we introduce some shorthand notation with 
\begin{equation} 
\begin{split}
\HnL{-1}^A_C  &\equiv  \ell_s^2\, \Kt^A_C \,,\\ 	 
\HnL{0} ^A_C&\equiv \frac{1}{2} \, m(m-1) \, \alpha_m \, \ell_s^{2m-2}\; \Kt_{A_1}^{C_1} \, \Kn_{B_1}^{D_1} \,
\left( {\cal R}^{m-2} \right)^{AA_1B_1}_{CC_1D_1} \,,\\
\HnL{n}^A_C  &= \partial_v^n \HnL{0}^A_C \,.
\label{}
\end{split}
\end{equation}	
Our task is to determine $\kappa_n$ so as to ensure that the resulting entropy function has the right monotonicity properties. 

Given this expression, we compute the temporal derivatives of $\sd_{m,\text{cor}}$ and combine it with the result in  Eq.~\eqref{eq:dvthetalove3} to obtain the answer for   $\Theta^{(m)} = \Theta_{m,\text{eq}} + \Theta_{m,\text{cor}}$. We will already write this in a compact form for further analysis:
 \begin{equation} \label{eq:dvthetalove4}
 \begin{split}
 \partial_v \Theta^{(m)} &= - T_{vv} - \Kt_{AC} \, \Kt^{AC}  + \nabla_A \mathcal{X}^A_{(m)} \\
 &\quad
	 - 4\, \ell_s^{-2} \, \HnL{1}^A_C \,  \HnL{-1}_{A}^{C}  
	 +4\, \tilde{\kappa}_0 \, \ell_s^{-2} \HnL{0}^2
	 \\
& \quad  	 
	+ \sum_{n=0}^\infty \, \kappa_n \, \ell_s^{2n}
	\Big\{
	2 \, \HnL{n+1}^2 + \HnL{n}^A_B \, \HnL{n+2}^B_A  \Big\} \\
& \quad  	 
	+ \sum_{n=0}^\infty \, \kappa_n \, \ell_s^{2n}	
	\Big\{ 2\, \Kt\, \HnL{n}^A_B\,\HnL{n+1}^B_A
	+ \partial_v \Kt\,  \HnL{n+1}^2 
	\Big\}
\end{split}
\end{equation}

The first term in the second line stems from rewriting Eq.~\eqref{eq:dvthetalove3} up to terms of order 
$\mathcal{O}(\alpha_m\Kt_{AC}^2 (\omega\ell_s)^p)$ for some $p\geq 0$ (c.f.\ the similar discussion around Eq.~\eqref{simpli}). We have again included a subleading term (proportional to $\tilde{\kappa}_0$) to put the final answer in a useful form. The terms in the last line may be ignored within our gradient expansion; the arguments for these is exactly as in the Gauss-Bonnet case, so we won't repeat ourselves here. All told, we can write the final expression in a compact form:
   \begin{equation} 
\begin{split}
  \partial_v \Theta^{(m)} &= - T_{vv} + \mathcal{J} +  \nabla_A \mathcal{X}^A_{(m)} \\
  \mathcal{J} &= 
   2 \sum_{n=-1}^\infty   \ell_s^{2n-2}
  	 \Big\{ \kappa_{n-1}\, \HnL{n}^2 +
  	 \kappa_n \, \ell_s^2\, \HnL{n}^A_B \, \HnL{n+2}^B_A  \Big\}  
 \end{split}
 \label{eq:LoveFinal} 	   
 \end{equation}
 Note that we have set specific values  $\kappa_{-2} = -\frac{1}{2}$, $\kappa_{-1}=-2$ and 
 $\tilde{\kappa}_0 =-1$  to bring the expression into the structure of a well-defined quadratic form.

 At this point we are done. Structurally, Eq.~\eqref{eq:LoveFinal} is identical to the analogous expression obtained for the Gauss-Bonnet theory, so the very same argument given in \S\ref{sec:scorrGB} will suffice 
 to demonstrate that the result is negative semidefinite. 

 Of course, we still have to tackle the boundary term involving the vector 
 $\mathcal{X}_{(m)}^A$  which is given in Eq.~\eqref{eq:LXterm}. Once again the comments made at the end of \S\ref{sec:GaussBonnet} continue to apply, owing to the occurrence of the $\Kt^C_A\, \nabla^{D_1} \Kt^{C_1}_{B_1}$ in the expression for $\mathcal{X}^{A}_{(m)}$ (with  appropriate contractions).  We however do not discern as yet a structure which ensures that this boundary contribution also remains negative, thereby preventing us from presenting a clean entropy function.

\subsection{Generalizations: Sums of Lovelock theories}
\label{sec:general}

If we view higher derivative theories of gravity as toy models for the subleading effects in a UV complete theory of quantum gravity, it is natural to consider not just isolated Lovelock terms for a given $m$. In this section we will argue that our analysis for isolated Lovelock terms can be generalized to this case with only minor modifications. The main insight will be a formula for the entropy functional in terms of the Lagrangian.

Let us consider an arbitrary Lovelock combination of the form 
\begin{equation}
\begin{split}
I &= \frac{1}{4\pi}\ \int d^dx\,  \sqrt{-g} \left({\cal L}_\text{grav} + \mathcal{L}_{matter} \right)\\
&\, {\cal L}_\text{grav} \equiv R + \sum_{m=2}^\infty \, \alpha_m\, 
\ell_s^{2m-2} \, \mathcal{L}_m \,,\\
&\qquad\; \equiv R + \sum_{m=2}^\infty \, \alpha_m\, 
\ell_s^{2m-2} \,
\delta^{\mu_1\nu_1\cdots \mu_m \nu_m}_{\rho_1\sigma_1\cdots \rho_m \sigma_m} \  R^{\rho_1}{}_{\mu_1}{}^{\sigma_1}{}_{\nu_1} \ \cdots\  R^{\rho_m}{}_{\mu_m}{}^{\sigma_m}{}_{\nu_m} \,.
\end{split}
\label{eq:action2}
\end{equation}

We start with the observation that the objects appearing in the proposed Lovelock entropy functional Eq.~\eqref{eq:LovelockEntropy} can be written in terms of derivatives of the corresponding Lagrangian: 
\begin{equation}
\frac{\delta {\cal L}_{m}}{\delta R^v{}_v{}^r{}_r} =m \, \alpha_m \; \ell_s^{2m-2} \;  \delta^{A_1B_1\cdots A_{m-1} B_{m-1}}_{C_1D_1 \cdots C_{m-1} D_{m-1}} \,  {\cal R}^{C_1}{}_{A_1}{}^{D_1}{}_{B_1} \cdots {\cal R}^{C_{m-1}}{}_{A_{m-1}}{}^{D_{m-1}}{}_{B_{m-1}} \,.
\end{equation}	
Define also the general expression for $\HnL{0}$ which is our basic building block:
\begin{equation}
\HnL{0}^A_B \equiv 
-  \frac{1}{2}\, \alpha_m \ell_s^{2m-2}  \; 
\frac{\delta^2 {\cal L}_\text{grav}}{
 \delta R^A{}_{A_1}{}^{C_1}{}_v \, \delta R^{v}{}_{B_1}{}^{D_1}{}_{B} } \bigg{|}_{R\rightarrow {\cal R}} \, \Kt_{A_1}^{C_1}\Kn_{B_1}^{D_1}
\label{}
\end{equation}	

We can use this observation to suggest a natural entropy functional for theories which involve several Lovelock terms:
\be \label{eq:Sfinal}
\begin{split}
S_\text{total} &= \int_{\Sigma_v} \sqrt{h} \; \Bigg\{1+ \sum_{m=2}^\infty \alpha_m\, \ell_s^{2m-2} \frac{\delta {\cal L}_{m}}{\delta R^v{}_v{}^r{}_r} \bigg{|}_{R\rightarrow {\cal R}} \\
&\qquad  + \sum_{n=0}^\infty \kappa_n \left[ \ell_s^{n}\, \partial_v^n \left( 
\sum_{m=2}^\infty \frac{1}{2}\alpha_m \ell_s^{2m-2}  \frac{\delta^2 {\cal L}_m}{\delta R^A{}_{A_1}{}^{C_1}{}_v \, \delta R^{v}{}_{B_1}{}^{D_1}{}_{B} } \bigg{|}_{R\rightarrow {\cal R}}  \, \Kt_{A_1}^{C_1}\Kn_{B_1}^{D_1}
\right) \right]^2
\Bigg\} \,.
\end{split}
\ee 
In this expression, we recognize the first line as Wald entropy. 
Since our analysis in  \S\ref{sec:lovelock} was linear in its use of the equations of motion, we would expect that the Wald entropy term in the first line (which is linear in ${\cal L}_m$) is reasonable.  
The second line provides the corrections in our perturbative framework, now expressed directly in terms of the Lagrangian. More succinctly, we can rewrite Eq.~\eqref{eq:Sfinal} as
\begin{equation}
\begin{split}
 S_\text{total} &= \int_{\Sigma_v} \sqrt{h} \, \left\{ \frac{\delta {\cal L}_\text{grav}}{\delta R^v{}_v{}^r{}_r} \bigg{|}_{R\rightarrow {\cal R}}
+ \sum_{n=0}^\infty \kappa_n
 \left[ \ell_s^n \partial_v^n \left( \frac{1}{2}\, \frac{\delta^2 {\cal L}_\text{grav}}{\delta R^A{}_{A_1}{}^{C_1}{}_v \, \delta R^{v}{}_{B_1}{}^{D_1}{}_{B} } \bigg{|}_{R\rightarrow {\cal R}}  \, \Kt_{A_1}^{C_1}\Kn_{B_1}^{D_1}
\right) \right]^2 \right\} 
\end{split}
 \label{eq:Sfinal2}
 \end{equation}
This equation should be viewed as the most compact and general result in our analysis. 

Let us now understand more explicitly why $S_\text{total}$ as defined in Eq.~\eqref{eq:Sfinal2} satisfies the second law. By superposing Lovelock terms from our previous analysis, we find that Eq.~\eqref{eq:Sfinal2} leads to the following:
 \begin{equation} \label{eq:dvthetalove4Mod}
 \begin{split}
\partial_v \Theta
 & = - T_{vv}  -\Kt^A_C \Kt^C_A  +  \nabla_A \mathcal{X}^A_{_\text{sum}}
\\
& \qquad -  4\, \ell_s^{-2}\, \Hntot{1}^A_C\Hntot{-1}^C_A
+ 4\, \tilde{\kappa}_0 \Hntot{0}^2 \\
&\qquad +  \sum_{n=0}^\infty \, \kappa_n\,\ell_s^{2n} \left\{2\, \Hntot{n+1}^2 +\Hntot{n}^A_C\,  \Hntot{n+2}^C_A\right\}  \\
&\qquad+\sum_{n=0}^\infty \kappa_n \, \ell_s^{2n}\,  \Big\{
 2\,\Kt\Hntot{n}^A_C \, \Hntot{n+1}^C_A 
 + \partial_v \Kt   \Hntot{n}^2 \Big\}  \\
&
\qquad    + \mathcal{O}(\alpha  \, \Kt_{AC}^2 (\omega\ell_s)^p)\,,
\end{split}
\end{equation}
where $p \geq 0$.  We define here the tensors:
\be\label{eq:Hntotdef}
\begin{split}
\Hntot{-1}^A_C \equiv \ell_s^2\, \Kt^A_C \,,\qquad 
 \Hntot{n\geq 0}^A_C \equiv \sum_{m=2}^\infty \,\HnL{n}^A_C \,.
\end{split}
\ee
The boundary term can also be understood in terms of the Lagrangian: 
{\footnotesize
\begin{equation}
\begin{split}
\mathcal{X}^{A_1}_{_\text{sum}} &= - \sum_{m=2}^\infty 2\,m(m-1) \alpha_m\, \ell_s^{2m-2}\, \left( {\cal R}^{m-3} \right)^{AA_1B_1A_2B_2}_{CC_1D_1C_2D_2}  \, \Kt_A^C \, \left( \nabla^{D_1} \Kt_{B_1}^{C_1} \right)  \\
&\hspace{3in}\times \left( {\cal R}^{C_2}{}_{A_2}{}^{D_2}{}_{B_2} - (m-2) \, \Kn_{A_2}^{D_2} \, \Kt_{B_2}^{C_2} \right)  \\
&= \left[ \frac{\delta^2 {\cal L}_\text{grav}}{\delta R^A{}_{A_1}{}^{C_1}{}_v \, \delta R^{v}{}_{B_1}{}^{D_1}{}_{B} }  -  \frac{\delta^3 {\cal L}_\text{grav}}{\delta R^A{}_{A_1}{}^{C_1}{}_v \, \delta R^{v}{}_{B_1}{}^{D_1}{}_{B} \, \delta R^{C_2}{}_{A_2}{}^{D_2}{}_{B_2}} \, \Kn_{A_2}^{D_2} \, \Kt_{B_2}^{C_2}\right]_{R\rightarrow {\cal R}} \Kt_A^B \, \left( \nabla^{D_1} \Kt_{B_1}^{C_1} \right) 
\end{split}
\label{eq:Xobsgen}
\end{equation}
}

As far as the second law is concerned, it is immediately clear that the expression Eq.~\eqref{eq:dvthetalove4Mod} can be assembled into a sum of squares with negative coefficients (up to negligible terms) in exactly the same way as we did in the case of a single Gauss-Bonnet or Lovelock term. The only difference lies in the more complicated building blocks Eq.~\eqref{eq:Hntotdef}. 

With regard to the obstruction term, one notes a general pattern, but it has not yet proven illuminating to see a useful bound. What remains clear is that one should perhaps try to control the tensor $\Kt_A^B \, \left( \nabla^{D_1} \Kt_{B_1}^{C_1} \right) $  on $\Sigma_v$. Modulo this issue, we have exhibited as before, a general entropy function for an arbitrary Lovelock theory.

\section{Discussion}
\label{sec:discussion}
 
We have constructed one entropy function for Lovelock theories of gravity. This entropy function  by construction always increases under every such time evolutions where the horizon remains spherically symmetric. 

It is possible to drop the assumption of spherical symmetry, provided we can control the total derivative obstruction term,  requiring it to be negative semidefinite. While we do not, as yet, have the general solution to this constraint, we think it is plausible that one can indeed construct such an entropy function. We now turn to describing a few subtleties regarding our construction, and also sketch out some potential extensions of the analysis herein.

\paragraph{Metric redefinition:}

If we take low energy limit of any UV complete theory of gravity, the classical action will generically have higher derivative corrections to all orders in the $\ell_s$ expansion. As  is well known  there exists a freedom of metric redefinition by terms of order  ${\cal O}(\ell_s^n),~n>0$, which will rearrange the form of the action. For example, we know that by a metric redefinition of the form
$$g_{\mu\nu}\rightarrow\tilde g_{\mu\nu} = g_{\mu\nu} +  \ell_s^2\left(\tilde a_1 R_{\mu\nu} + \tilde a_2 R~ g_{\mu\nu}\right)$$
we can transform any four derivative pure gravity action to Gauss-Bonnet. This, in turn, implies that in a gravity action, where terms of all order in ${\cal O}(\ell_s^2)$ are allowed, at four-derivative order there could at most be one free coefficient which is of physical significance; the remaining two could always be absorbed in an appropriate field redefinition (in the absence of matter).\footnote{ However, this metric-redefinition freedom does not exist,  if we demand that our gravity action truncates at four-derivative order which, one suspects, does not give rise to a consistent quantum theory \cite{Camanho:2014apa}.} 

The situation is less clear at higher orders in $\ell_s$; at six-derivative order it appears that by field redefinitions one can reduce the number of independent data to two functional forms. One of these is the Lovelock term $\mathcal{L}_3$, but the other is a quasi-topological form involving a different index contraction of curvatures \cite{Oliva:2010eb,Myers:2010ru,Oliva:2011xu}. It is clear from this analysis that with increasing derivative order, more tensor structures are possible (see \cite{Deser:2016tgn} for related comments). So within effective field theory, we should a-priori allow other possible forms of the gravitational interactions, differing from the Lovelock terms considered herein.

Now, under an arbitrary metric redefinition, a null hypersurface embedded in a dynamical spacetime might not remain null. Our analysis heavily used the fact that the horizon is a null surface throughout the time evolution. In other words, we have chosen to fix our field redefinitions so as to ensure that the metric which appears in the action is the very same one that imposes the causal structure on the spacetime. We require the horizon to be a null hypersurface with respect to this choice. It is an interesting open question, as to how to properly  assign an entropy functional to an action, without a-priori fixing a field redefinition frame.  Such a viewpoint would be really useful in ascertaining an entropy function for arbitrary higher derivative interactions, an ambitious task whose surface we  have barely scratched. 

\paragraph{Coordinate choice:}

Throughout our analysis we have chosen a very special gauge adapted to the horizon. The spacetime is foliated by constant $r$ slices such that the event horizon is at $r=0$. But our choice does not fix the coordinate transformation freedom completely. For example, a simple constant scaling (see \cite{Wall:2015raa}) of $v\rightarrow\tilde v =\lambda~ v$ and $r\rightarrow\tilde r= \frac{r}{\lambda}$, coupled with a change $f(r,v)\rightarrow \tilde f(\tilde r,\tilde v)= \frac{f(r,v)}{\lambda^2}$  keeps the form of the metric invariant (see Eq.~\eqref{metricintro}). In particular, since $f(r=0,v)=0$, on horizon one does not even need to change any of the metric components and the metric is exactly invariant at $r=0$.  However, the corrected entropy density, though defined only on the horizon,  is not automatically invariant under this transformation unless we also scale the $\kappa_n$'s appropriately.

It would be interesting to classify all such transformations that keep the form of the metric invariant and to see how the formula of $\partial_v\Theta$ changes as we change the foliation of the spacetime.

Another potential drawback is that we have had to pick a particular foliation of the 
horizon, with leaves given by $\Sigma_v$, and thence construct a scalar function on these leaves. A more covariant procedure, suggested by the fluid/gravity analogy, would have been to construct directly a spacetime codimension-2 form, which could then be integrated on any arbitrary spatial section of $\mathcal{H}^+$. Rather curiously, we were unable to construct such an entropy-form (whose dual would be the entropy current). While we do not have a formal no-go result, indications are that such a covariant object does not exist, a fact that we find rather surprising.\footnote{ For instance, using the fluid/gravity map in asymptotically AdS spacetimes \cite{Hubeny:2011hd}, we could have pushed forward the entropy current onto the horizon. We thank Shiraz Minwalla for numerous discussions regarding this issue. }

\paragraph{Total derivative terms:}

We have seen that if we depart from spherical symmetry, then for Lovelock theories our entropy function works provided a very particular term, which is of the form of a total derivative, is either vanishing or always non-negative. Note that though we have assumed that our horizon is a compact hypersurface and we are concerned only with the total entropy, this particular total derivative term cannot be removed since it appears in the  $v$ derivative of the integrand of $\partial_v S_{\text{total}}$ (see Eq.~\eqref{reldlth1a}). Note also that this particular term occurs at the same order in the gradient expansion where the higher derivative terms first start contributing to the rate of entropy change.

It would be very interesting to  know whether this term has any physical significance and says something important about the theories of gravity. It is also possible that by doing some re-adjustment in our construction we could absorb this term also in a sum of full square pieces. Perhaps some redefinition in the foliation of the spacetime (as mentioned in the previous paragraph) along with the addition of another infinite series in the expression of entropy will suffice to construct a function with a negative semidefinite second derivative.

\paragraph{Other possible methods:}

Another possibility is that instead of proving $\partial_v \Theta \leq 0$, we could try to prove $\partial_v (Z\Theta) \leq 0$ with $Z$ being a positive function. As we review in Appendix
\ref{sec:fR}, the standard proof of second law in $f(R)$ theories by \cite{Jacobson:1995uq} proceeds exactly by this method. More generally, while we have found an obstruction to
proving second law using usual Raychaudhuri equation, our work does not preclude that there might be some other way of proving second law. We hope our work will be a stepping stone
and an inspiration for an eventual proof.

\paragraph{Future directions:} Other future directions include to examine our results in the regime of fluid/gravity
\cite{Hubeny:2011hd}, or in the context of large $D$ black holes dual to membrane dynamics  \cite{Bhattacharyya:2016nhn}. In particular, it would be interesting to examine the total derivative obstruction  term in these contexts to see whether we can actually use it to engineer entropy destruction in generic solutions.  We leave  these studies for future work.

\acknowledgments

It is a great pleasure to thank Shiraz Minwalla for initiating discussions on this topic,  for an extended collaboration, and for his numerous suggestions and enlightening comments throughout the course of this work.  
We would also like to thank  Arpan Bhattacharyya, Jyotirmoy Bhattacharya, Srijit Bhattacharjee, Joan Camps, Veronika Hubeny, Dileep Jatkar, Juan Maldacena, 
Sudipta Sarkar, Ashoke Sen, Aninda Sinha,  and Tadashi Takayanagi for illuminating discussions.
FH gratefully acknowledges support through a fellowship by the Simons Foundation.
MR would like to thank IAS and ICTS for hospitality during the course of this project. Research of N.K. is supported by the JSPS Grant-in-Aid for Scientific Research (A) No.16H02182. N.K. acknowledges the support he received from his previous affiliation HRI, Allahabad. N.K. would also like to thank IITK for hospitality during the course of this project. The work of S.B. was supported by an India Israel (ISF/UGC) joint research grant (UGC/PHY/2014236).  S.B. and R.L.  would  also like to acknowledge our debt to the people of India for their steady and generous support to research in the basic sciences.


\appendix

\section{Coordinate choice and near-horizon metric} 
\label{appC}

We argue here that the metric used in the main text Eq.~\eqref{eq:metric} is the most general such, justifying the statements made in \S\ref{sec:geom}. 

Let us w.l.o.g. start with the metric in Eq.~\eqref{metch2}. In writing this we have only assumed that we have a spacetime codimension-1 null hypersurface, foliated by spacetime codimension-2 slices $\Sigma_v$ whose null normals are $t^\mu$ and $n^\mu$ coordinatized as explained earlier.
The inverse metric components are
\be \label{invgintmed}
\begin{split}
&g^{rr} = \frac{f + k^2}{\Delta},~~ g^{rv} = \frac{j -( J\cdot k)}{\Delta},~~g^{vv} = \frac{J^2}{\Delta}\\
&g^{rA} = - \left(g^{rr} J^A+g^{rv} k^A\right),~~~g^{vA} = - \left(g^{rv} J^A+g^{vv} k^A\right)\\
&g^{AB} = h^{AB} + g^{rr} J^A J^B + g^{vv} k^A k^B + g^{rv} \left(J^A k^B + k^A J^B\right)
\end{split}
\ee
where $~\Delta = \left[j-(J\cdot k)\right]^2 - J^2 (f + k^2)$ and $h^{AB} $ is the inverse metric on 
$\Sigma_v$. All indices are raised/lowered as everywhere else in the text with the metric $h_{AB}$, e.g.,  $J^A= h^{AB}J_B$. Also, we have used the following definitions  
\be
J\cdot k \equiv h^{AB} J_A k_B,~~ J^2\equiv h^{AB}J_A J_B,~~ k^2\equiv h^{AB}k_A k_B.
\ee

Consider the vector $\partial_r = n^{\mu}\partial_\mu$, such that $n^r=1\,,~n^vni^A=0$. In order to make it an affinely parametrized null generator we need to impose
\be
n^{\mu} D_{\mu} n^{\nu} = 0\,,
\ee
which is evaluated to be
\be \label{parvaffine}
\begin{split}
n^{\mu} \nabla_{\mu} n^{r} &= \Gamma^r_{rr}=g^{rv} \partial_r J(r,v,{\bf x}) + g^{rA} \partial_r J_A(r,v,{\bf x})\,, \\
n^{\mu} \nabla_{\mu} n^{v} &= \Gamma^v_{rr}=g^{vv} \partial_r J(r,v,{\bf x}) + g^{vA} \partial_r J_A(r,v,{\bf x})\,, \\
n^{\mu} \nabla_{\mu} n^{A} &= \Gamma^A_{rr}=g^{Av} \partial_r J(r,v,{\bf x}) + g^{AB} \partial_r J_B(r,v,{\bf x}).
\end{split}
\ee
Using Eq.~\eqref{parvaffine} and the expressions of inverse metric as given in Eq.~\eqref{invgintmed} it is easy to verify that 
\be 
J^A ~\Gamma^r_{rr} + k^A  ~\Gamma^v_{rr} +  \Gamma^A_{rr} = h^{AB} \partial_r J_B(r,v,{\bf x}) =0
\ee
which immediately reduces to the condition 
\be \label{cc1}
\partial_r J_B(r,v,{\bf x}) =0 \,.
\ee
Plugging this  back in Eq.~\eqref{parvaffine} it is straightforward to obtain 
\be \label{cc2}
\partial_r j(r,v,{\bf x}) =0.
\ee
From  Eq.~\eqref{cc1} and  Eq.~\eqref{cc2} it follows that $j(r,v,x)$ and $J_A(r,v,x)$ are actually independent of $r$ and functions of $v$ and $x^A$ alone.
\be \label{cc}
\begin{split}
 j(r,v,{\bf x}) &= \phi(v,{\bf x})\\
J_A(r,v,{\bf x})&=\Phi_A(v,{\bf x})
\end{split}
\ee

Now  the third line of  Eq.~\eqref{metch2} specifies the angle at which the $\partial_r$ geodesics are piercing through the horizon ($r=0$), and thereby sets the initial condition  Eq.~\eqref{cc}. It simply implies that
\begin{equation}\label{sts}
\begin{split}
&\phi(v,{\bf x}) = j(r=0,v,{\bf x}) =1,~~~~~~~\text{for all}~\{v,{\bf x}\}\\
&\Phi_A(v,{\bf x}) = J_A(r=0,v,{\bf x}) =0,~~~\text{for all}~\{v,{\bf x}\}
\end{split}
\end{equation}
This is the statement of Eq.~\eqref{parraffcond}.

Before we move to discussing the constraint that follows from demanding the other null vector $\partial_v= t^{\mu}\partial_\mu$ (with $t^v=1$ and $t^r=t^A=0$), let us write the metric incorporating the constraints in Eq.~\eqref{parraffcond},
 \begin{equation}\label{metch2app}
 \begin{split}
 &ds^2 = 2 dv ~dr-f(r,v,x) dv^2 + 2\,k_A (r,v,x) dv dx^A+ h_{AB}(r,v,x) dx^A dx^B\, ,\\
 &\text{such that} ~~f(r=0,v,x) =0,~k_A(r=0,v,x) =0\, .
 \end{split}
 \end{equation}
The fact that $t^{\mu}$ is an affinely parametrized null generator requires that $t^{\mu} D_{\mu} t^{\nu} = 0$, when evaluated on the horizon. This condition reduces to
\be
t^{\mu} D_{\mu} t^{v} |_{r=0} = \Gamma^v_{vv} = {1\over 2} \partial_r f |_{r=0}
\ee
Therefore, demanding that $\partial_v$ is an affinely parametrized null generator on the horizon leads us to the constraint that $\partial_r f |_{r=0}=0$. Hence, we have now justified our claims in \S\ref{sec:geom}.

\section{Curvature tensors for stationary metric}
\label{appallcurv}

In this appendix we will first work out the curvature tensors for the metric 
Eq.~\eqref{metch2app}. Its inverse is easily obtained:
\begin{equation}\label{inversemetric}
\begin{split}
&g^{rr} = f + k^2 ,~~g^{rv} = 1,~~g^{rA} = -k^A \,,\\
&g^{vv} = 0,~~g^{vA}= 0,~~g^{AB} = h^{AB}\,,
\end{split}
\end{equation}
with $$h^{AB} = (h^{-1})_{AB},~~k^A= h^{AB} k_B,~~ k^2 = h^{AB}k_A k_B\,.$$

The Christoffel symbols  are also easily computed to be:
\begin{equation}\label{christoffel}
\begin{split}
&\Gamma^{\mu}_{rr} =0,~~~~\Gamma^v_{r \mu} =0\\
&\Gamma^r_{vr} = -\frac{1}{2}\partial_r f - \frac{k^A}{2}\partial_r k_A,~~\Gamma^r_{Ar} = \frac{1}{2} \partial_r k_A -\frac{1}{2}k^B\partial_rh_{AB}\\
&\Gamma^r_{vv} = \frac{1}{2}(f + k^2) \partial_r f -\frac{1}{2}\partial_v f - k^A\left(\partial_v k_A + \frac{1}{2}\partial_A f\right)\\
&\Gamma^r_{AB} =-k_C\hat\Gamma^C_{AB} -\frac{1}{2} (f + k^2) \partial_r h_{AB} +\frac{1}{2}\left(\partial_A k_B + \partial_B k_A -\partial_v h_{AB}\right)\\
&\Gamma^v_{vv} = \frac{1}{2}\partial_r f ,~~~\Gamma^v_{vA} = -\frac{1}{2}\partial_r k_A ,~~~\Gamma^v_{AB} = -\frac{1}{2}\partial_r h_{AB}\\
&\Gamma^A_{rv}= \frac{1}{2} h^{AB} \partial_r k_B,~~\Gamma^A_{rB} = \frac{1}{2} h^{AC} \partial_r h_{BC}\\
&\Gamma^A_{vv} = -\frac{k^A}{2}\partial_r f + h^{AB}\left(\partial_v k_B + \frac{1}{2}\partial_B f\right)\\
&\Gamma^A_{vB} = \frac{k^A}{2}\partial_r k_B - \frac{h^{AC}}{2}\left[\partial_C k_B - \partial_B k_C -\partial_v h_{BC}\right]\\
&\Gamma^A_{BC} = \frac{k^A}{2} \partial_r h_{BC} + \hat\Gamma^A_{BC}
\end{split}
\end{equation}
where $\hat\Gamma^A_{BC}$ is the Christoffel symbol for the spatial metric $h_{AB}$. 

Before we proceed let us record some more notation:
\begin{equation}\label{notation2}
\begin{split}
\partial_r h_{AB} = 2\,  \Kn_{AB},~~\partial_v h_{AB} = 2\, \Kt_{AB},~~ \partial_r k_A = 2\, \omega_A,~~\partial_r f = 2\,T.
\end{split}
\end{equation}
Let us also use Eq.~\eqref{metricintroa} to record the non-zero Christoffel symbols on the horizon 
 $\mathcal{H}^+$ 
\begin{equation}\label{christhorizon}
\begin{split}
&\Gamma^r_{Ar}\big|_{\mathcal{H}^+}=\omega_A\,, ~~~~\Gamma^r_{AB}\Big|_{\mathcal{H}^+}= - \Kt_{AB}\,,~~~~\Gamma^v_{Av}\Big|_{\mathcal{H}^+}=-\omega_A\,, ~~~~\Gamma^v_{AB}\Big|_{\mathcal{H}^+}= -\Kn_{AB}\\
&\Gamma^A_{rv}\Big|_{\mathcal{H}^+} = \omega^A\,,~~~~\Gamma^A_{rB}\Big|_{\mathcal{H}^+}=\Kn^A_B\,, ~~~~\Gamma^A_{vB}\Big|_{\mathcal{H}^+}= \Kt^A_B\,,~~\Gamma^A_{BC}\Big|_{\mathcal{H}^+}= \hat\Gamma^A_{BC} \,.\\
\end{split}
\end{equation}
Using these we find the non-zero components of the Riemann curvature tensor to be given by the following:
\be \label{riemannapp}
\begin{split}
R_{rvrv} &= {1 \over 2} \partial_r^2 f + \omega^2\\
R_{rvrA} &= -\partial_r \omega_A+\omega^B \Kn_{AB}\\
R_{rvvA} &= -\partial_v \omega_A-\omega^B \Kt_{AB}\\
R_{rArB} &= -\partial_r \Kn_{AB}+ \Kn_{AC}\Kn^C_B\\
R_{rAvB} &= -\partial_r \Kt_{AB} + \nabla_B \omega_A - \omega_A\omega_B + \Kn_{BC} \Kt^{C}_A\\
R_{vAvB} &= -\partial_v \Kt_{AB}+ \Kt_{AC}\Kt^C_B\\
R_{ABvC} &= \nabla_B \Kt_{AC} -  \nabla_A \Kt_{BC}\\
R_{ABrC} &= (\nabla_B - \omega_B) \Kn_{AC} - (\nabla_A - \omega_A) \Kn_{BC}\\
R_{ABCD} &=\mathcal{R}_{ABCD} + \Kn_{AC}\Kt_{BD} + \Kt_{AC}\Kn_{BD}- \Kn_{AD}\Kt_{BC} - \Kt_{AD}\Kn_{BC}
\end{split}
\ee
where $\nabla_A$ is the covariant derivative with respect to the metric $h_{AB}$.
Furthermore, the Ricci tensor evaluated at the horizon takes the form:
\be
\begin{split}
R_{rr} &= -\partial_r \Kn - \Kn_{AB}\Kn^{AB}\\
R_{rv} &= {1 \over 2} \partial_r^2 f- \partial_r \Kt -  \Kn_{AB}\Kt^{AB} + \nabla^A \omega_A \\
R_{rA} &= \partial_v \omega_A + \Kn_A^B \omega_B + (\nabla_B - \omega_B)\Kn^B_A - (\nabla_A - \omega_A) \Kn \\
R_{vv} & = -\partial_v \Kt - \Kt_{AB}\Kt^{AB} \\
R_{vA} &= -\partial_v \omega_A + \Kt^B_A \omega_B +\nabla_B \Kt^B_A -\nabla_A \Kt\\
R_{AB} &=\mathcal{R}_{AB} - 2 \partial_r \Kt_{AB} +(\nabla_B \omega_A +\nabla_A \omega_B - 2 \omega_A\omega_B)\\ & ~~~~~~~~ +2 (\Kn_{AC}\Kt^C_B + \Kn_{BC}\Kt^C_A)  - \Kn \Kt_{AB} -\Kt \Kn_{AB}
\end{split}
\ee
 Note that the trace of $\Kt_{AB}$ and $\Kn_{AB}$ are denoted as $\Kt$ and $\Kn$,  respectively $\Kt = \Kt_{AB} h^{AB},~~ \Kn = \Kn_{AB} h^{AB}$.

\paragraph{Derivation of Eq.~\eqref{intcurid}:}
With the expressions for the curvature tensors, as in Eq.~\eqref{riemannapp}, at our disposal we shall now give the procedure to derive Eq.~\eqref{intcurid}. To this end let us focus on the last equation in the set of Eq.~\eqref{riemannapp} and rewrite it in terms of the codimension-2 Gauss-Coddazi equation:
\be \label{riemannapp1}
\begin{split}
{{{R^{C_1}}_{A_1}}^{D_1}}_{B_1} - {{{\mathcal{R}^{C_1}}_{A_1}}^{D_1}}_{B_1}  &= \Kn_{A_1 B_1} \Kt^{C_1 D_1} + \Kt_{A_1 B_1} \Kn^{C_1 D_1} - \Kn^{D_1}_{A_1} \Kt^{C_1}_{B_1} - \Kn^{C_1}_{B_1} \Kt^{D_1}_{A_1} \,.
\end{split}
\ee
From Eq.~\eqref{riemannapp1} it is straightforward to conclude that
\be\label{eq:deltaR}
\begin{split}
&\delta^{A_1A_2B_2}_{D_1C_2D_2}\bigg[{{{R^{C_2}}_{A_2}}^{D_2}}_{B_2} - {{{\mathcal{R}^{C_2}}_{A_2}}^{D_2}}_{B_2} \bigg] =
-2 \delta^{A_1A_2B_2}_{D_1C_2D_2} \Kn^{D_2}_{A_2} \Kt^{C_2}_{B_2} \,,
\end{split}
\ee
where $\delta^{A_1A_2B_2}_{D_1C_2D_2}$ is the now familiar generalized Kronecker symbol (see \S\ref{sec:summaryresult}). This then leads to the expression  Eq.~\eqref{intcurid} which we used in the text.

\section{Computational details: Gauss-Bonnet theory}
\label{appdtlcalc}

In this appendix we shall fill in some of the intermediate steps for the calculations described in  \S\ref{sec:GaussBonnet}. 

\subsection{Derivation of $\Theta_{2,\text{eq}}$}
 \label{appA2}
 
We start with $S_\text{Wald}$ as given in Eq.~\eqref{enteq} and then compute

\begin{equation}
\begin{split}
\partial_v S_\text{Wald}&= \partial_v  \left[\int_{\Sigma_v} d^{d-2}x \, \sqrt{h} \;(1+ \alpha_2\,\ell_s^2\;  \sd_{2,\text{eq}}) \right] \\ 
&= \int_{\Sigma_v} d^{d-2}x \,  \left[\Kt( 1 + \alpha_2\,\ell_s^2\;  \sd_{2,\text{eq}}) + \alpha_2\,\ell_s^2\;  \partial_v \sd_{2,\text{eq}}\right]
\\
&=\int_{\Sigma_v} d^{d-2}x \, \left[\Kt( 1 + \alpha_2\,\ell_s^2\;  \sd_{2,\text{eq}}) - 2\alpha_2\,\ell_s^2\; \left(\frac{\delta \sd_{2,\text{eq}}}{\delta h^{AB}}\right)\Kt^{AB}\right] .
\end{split}
\end{equation}
and hence from Eq.~\eqref{deftheta} we get
\begin{equation} \label{thetaeqapp}
\begin{split}
\Theta_{2,\text{eq}} = \Kt( 1 + \alpha^2 \sd_{2,\text{eq}}) - 2\alpha^2\left(\frac{\delta \sd_{2,\text{eq}}}{\delta h^{AB}}\right)\Kt^{AB}\,.
\end{split}
\end{equation}
Using the expression of $\sd_{2,\text{eq}}$ from Eq.~\eqref{defrho}, and the variational identity 
Eq.~\eqref{drdh1}, we arrive at 
\be \label{drdh2a}
{\partial \sd_{2,\text{eq}} \over \partial h^{AB} }= 2\delta^{A_1 B_1 }_{C_1 D_1} ~{\partial h^{D_1 \tilde{D_1}} \over \partial h^{AB} } \mathcal{R}^{C_1}_{A_1 \tilde{D_1} B_1} = -\delta^{D A_1 B_1}_{B C_1 D_1}~ {{{\mathcal{R}^{C_1}}_{A_1}}^{D_1}}_{B_1}h_{AD}\,,
\end{equation}
In writing this expression we have dropped the total derivative term where we vary the curvature tensor. This  last equality has been simplified using the following relations:
\begin{equation}
\begin{split}
{\partial h^{D_1 \tilde{D_1}} \over \partial h^{AB} } = \delta^{D_1}_{A}  \delta^{\tilde{D_1}}_{B},~~
\delta^{AA_1B_1} _{CC_1D_1} &=\delta^{A}_{C} \delta^{A_1B_1} _{C_1D_1} -\delta^{A}_{C_1} \delta^{A_1 B_1} _{C D_1}+\delta^{A}_{D_1} \delta^{A_1B_1} _{C C_1}\,.
\end{split}
\end{equation}
Finally, using Eq.~\eqref{drdh2a} back in Eq.~\eqref{thetaeqapp} we derive
Eq.~\eqref{eqtheta}.

\subsection{Gauss-Bonnet  equation of motion}
\label{appA3}

Let us now turn to the equation of motion for Gauss-Bonnet gravity. The action for the Gauss-Bonnet theory is given in Eq.~\eqref{GB1}. Varying the action with respect to the spacetime metric $g_{\mu\nu}$ we get the equations of motion
\begin{equation}
\begin{split}
R_{\mu \nu } -{1\over 2} g_{\mu \nu } R +\alpha_2\, \ell_s^2 \mathcal{E}_{\mu \nu}=  T_{\mu\nu}\,,
\end{split}
\end{equation}
where the matter stress tensor was defined in Eq.~\eqref{eq:Tdef}. We have included the contribution of the Gauss-Bonnet term in $\mathcal{E}_{\mu\nu}$ separately since it occurs at higher orders in our gradient expansion. This term is evaluated as follows:
\begin{equation}\label{Eom1}
\begin{split}
\mathcal{E}_{\mu \nu} &= \frac{\delta \mathcal{L}_2 }{\delta g^{\mu \nu }} - \frac{g_{\mu \nu }}{2}  \mathcal{L}_2 \\
&=2\delta^{\mu_1 \nu_1\mu_2 \nu_2}_{\rho_1 \sigma_1\rho_2 \sigma_2}\left[\frac{\delta{ R^{\rho_1}}_{\mu_1\alpha\nu_1}g^{\alpha \sigma_1}}{\delta g^{\mu \nu}}\right]{{{{R^{\rho_2}}_{\mu_2}}^{\sigma_2}}_{\nu_2}} - \frac{g_{\mu \nu }}{2}  \mathcal{L}_2\\
&=2\delta^{\mu_1 \nu_1\mu_2 \nu_2}_{\rho_1 \sigma_1\rho_2 \sigma_2}\bigg[{ R^{\rho_1}}_{\mu_1(\mu\nu_1}\delta^{\sigma_1}_{\nu)} + g^{\alpha\sigma_1}\left(D_{\alpha} \frac{\delta\Gamma^{\rho_1}_{\mu_1 \nu_1}}{\delta g^{\mu \nu}} -D_{\nu_1} \frac{\delta\Gamma^{\rho_1}_{\mu_1 \alpha}}{\delta g^{\mu \nu}}\right) \bigg]{{{{R^{\rho_2}}_{\mu_2}}^{\sigma_2}}_{\nu_2}} - \frac{g_{\mu \nu }}{2}  \mathcal{L}_2\,. \\
\end{split}
\end{equation}
In Eq.~\eqref{Eom1} the part that involves the variation of the Christoffel symbols can be manipulated to be written as total derivatives and hence will not appear in the equations of motion. We shall now demonstrate this explicitly
\begin{equation}\label{Eom2}
\begin{split}
&\delta^{\mu_1 \nu_1\mu_2 \nu_2}_{\rho_1 \sigma_1\rho_2 \sigma_2} ~g^{\alpha\sigma_1}\left(D_{\alpha} \delta\Gamma^{\rho_1}_{\mu_1 \nu_1} -D_{\nu_1} \Gamma^{\rho_1}_{\mu_1 \alpha}\right){{{{R^{\rho_2}}_{\mu_2}}^{\sigma_2}}_{\nu_2}}\\
&=~- \delta^{\mu_1 \nu_1\mu_2 \nu_2}_{\rho_1 \sigma_1\rho_2 \sigma_2} ~ g^{\alpha \sigma_1}\left(D_{\nu_1} \delta\Gamma^{\rho_1}_{\mu_1 \alpha}\right) {{{{R^{\rho_2}}_{\mu_2}}^{\sigma_2}}_{\nu_2}}\\
&=~ \delta^{\mu_1 \nu_1\mu_2 \nu_2}_{\rho_1 \sigma_1\rho_2 \sigma_2}~ g^{\alpha \sigma_1}  \delta\Gamma^{\rho_1}_{\mu_1 \alpha} ~ \left(D_{\nu_1}{{{{R^{\rho_2}}_{\mu_2}}^{\sigma_2}}_{\nu_2}}\right) \\
&=- \frac{1}{2}\delta^{\mu_1 \nu_1\mu_2 \nu_2}_{\rho_1 \sigma_1\rho_2 \sigma_2} \left[ D_{\mu_1} \delta g^{\rho_1 \sigma_1} - D^{\rho_1} \delta g ^{\sigma_1}_{\mu_1} + D^{\sigma_1} \delta g^{\rho_1}_{\mu_1} \right]\left(D_{\nu_1}{{{{R^{\rho_2}}_{\mu_2}}^{\sigma_2}}_{\nu_2}}\right) \\
&=-\delta^{\mu_1 \nu_1\mu_2 \nu_2}_{\rho_1 \sigma_1\rho_2 \sigma_2} \left[ D^{\sigma_1} \delta g^{\rho_1}_{\mu_1} \right](D_{\nu_1} {{{{R^{\rho_2}}_{\mu_2}}^{\sigma_2}}_{\nu_2}})\,.
\end{split}
\end{equation}
Note that in the second equality we have discarded a total derivative term, and in the third equality we have used the relation
\be
g^{\alpha \sigma_1}  \delta\Gamma^{\rho_1}_{\mu_1 \alpha} = -{1 \over 2} \bigg[ D_{\mu_1} \delta g^{\rho_1 \sigma_1} - D^{\rho_1} \delta g ^{\sigma_1}_{\mu_1} + D^{\sigma_1} \delta g^{\rho_1}_{\mu_1}\, \bigg].
\ee

We can use the Bianchi identity to show that $D_{[\nu_1} {{{{R^{[\rho_2}}_{\mu_2}}^{\sigma_2]}}_{\nu_2]}}$ vanishes:
\begin{equation}\label{bianchi1}
\begin{split}
D_{[\nu_1} {{{{R^{[\rho_2}}_{\mu_2}}^{\sigma_2]}}_{\nu_2]}}
&=-D_{[\nu_1} \left({R^{[\rho_2\sigma_2]}}_{\nu_2 \mu_2]} + {{R^{[\rho_2}}_{\nu_2 \mu_2]}}^{\sigma_2]}\right)\\
&=-D_{[\nu_1} \left( {{R^{[\rho_2}}_{\nu_2 \mu_2]}}^{\sigma_2]}\right)= -D_{[\nu_1} {{{{R^{[\rho_2}}_{\mu_2}}^{\sigma_2]}}_{\nu_2]}}=0\,.
\end{split}
\end{equation}
Therefore we conclude that
\begin{equation}\label{Eom3}
\begin{split}
\mathcal{E}^{\alpha}_{\beta} &= 2 \delta^{ \mu_1 \nu_1\mu_2\nu_2}_{ \rho_1 \beta \rho_2 \sigma_2}  {{{{R^{\rho_1}}_{\mu_1}}^{\alpha}}_{\nu_1}} {{{{R^{\rho_2}}_{\mu_2}}^{\sigma_2}}_{\nu_2}}
 - \frac{\delta^{\alpha}_{\beta}}{2}  \mathcal{L}_2 \\ 
 &=-\frac{1}{2}\delta^{\alpha \mu_1 \nu_1 \mu_2 \nu_2}_{\beta \rho_1 \sigma_1 \rho_2 \sigma_2}\left[{{{{R^{\rho_1}}_{\mu_1}}^{\sigma_1}}_{\nu_1}}  {{{{R^{\rho_2}}_{\mu_2}}^{\sigma_2}}_{\nu_2}}\right]\,.
  \end{split}
 \end{equation}
 In the last line we have used the fact that $\delta^{\mu_1 \nu_1\mu_2 \nu_2}_{\rho_1 \sigma_1\rho_2 \sigma_2}$ is a determinant and therefore could be expanded in terms of  cofactors as written below
 $$\delta^{\mu_1 \nu_1\mu_2 \nu_2}_{\rho_1 \sigma_1\rho_2 \sigma_2} = \delta^{\mu_1}_{\rho_1} \delta^{\nu_1\mu_2 \nu_2} _{\sigma_1\rho_2\sigma_2} -\delta^{\mu_1}_{\sigma_1} \delta^{\nu_1\mu_2 \nu_2} _{\rho_1\rho_2\sigma_2}  + \delta^{\mu_1}_{\rho_2} \delta^{\nu_1\mu_2 \nu_2} _{\rho_1 \sigma_1\sigma_2}-\delta^{\mu_1}_{\sigma_2} \delta^{\nu_1\mu_2 \nu_2} _{\rho_1 \sigma_1\rho_2}.$$
Projecting onto the horizon $\mathcal{H}^+$ in  the geometry Eq.~\eqref{metricintro}, we obtain the temporal component that we need:
\begin{equation}
\label{gbeom2}
\begin{split}
\mathcal{E}_{vv}&= 2(R R_{vv} -2R_{v\alpha}R^{\alpha}_v - 2 R^{\alpha \beta}R_{v\alpha v \beta}+{R_v}^{\alpha\beta\rho}R_{v\alpha\beta\rho}) \\ & =  -2~ \delta^{A_1 A_2 B_2}_{D_1 C_2 D_2} \left[\left(\frac{d \Kt^{D_1}_{A_1} }{d v} + \Kt_{A_1C}\Kt^{D_1C} \right) {{{{R^{C_2}}_{A_2}}^{D_2}}_{B_2}}-2  \nabla_{A_2} \Kt^{D_1}_{A_1}~\nabla^{D_2}\Kt^{C_2}_{B_2}\right] \,.
\end{split}
\end{equation}
This form of the equation has been used in \S\ref{sec:GaussBonnet} to simplify Term 1 in Eq.~\eqref{reldlth}.
 
\subsection{Simplifying Wald entropy change} 
\label{appA4}

We provide some details regarding the derivation of Eq.~\eqref{cont3}. 
We are looking at the third term in Eq.~\eqref{reldlth}, labeled Term $3$, 
which is of the form 
$2 \alpha_2\, \ell_s^2\; \Kt_A^C \delta^{A A_1 B_1}_{C C_1 D_1}~\partial_{v} {{{\mathcal{R}^{C_1}}_{A_1}}^{D_1}}_{B_1}$.
In order to proceed we need to calculate derivative of the intrinsic Riemann tensor ${{{\mathcal{R}^{C_1}}_{A_1}}^{D_1}}_{B_1}$ with respect to $v$
\begin{equation}\label{inbetween1}
\begin{split}
&\partial_{v} {{{\mathcal{R}^{C_1}}_{[A_1}}^{D_1}}_{B_1]}= -h^{CD_1}\nabla_{[B_1}\partial_v \Gamma^{C_1}_{A_1]C} - 2 {{{{\mathcal{R}^{C_1}}_{[A_1}}^{D}}_{B_1]}}\Kt^{D_1}_D\\
&= -\nabla_{B_1} \left(\nabla_{A_1} \Kt^{C_1 D_1} + \nabla^{D_1} \Kt^{C_1}_{A_1} - \nabla^{C_1} \Kt^{D_1}_{A_1}\right)- 2 {{{{\mathcal{R}^{C_1}}_{[A_1}}^{D}}_{B_1]}}\Kt^{D_1}_D \,.
\end{split}
\end{equation}
Using Eq.~\eqref{inbetween1} and the property of $\delta^{A A_1 B_1}_{C C_1 D_1}$ that it is antisymmetric in both in its upper and lower indices, it is easy to obtain
\be  \label{cont3app}
\begin{split}
&\Kt_A^C \delta^{A A_1 B_1}_{C C_1 D_1}~\partial_{v} {{{\mathcal{R}^{C_1}}_{A_1}}^{D_1}}_{B_1} =
2~\delta^{A_1 A_2 B_2}_{D_1 C_2 D_2} \nabla_{A_2} \big( \Kt^{D_1}_{A_1}\nabla^{D_2}\Kt^{C_2}_{B_2} \big)\\
&\quad -2~\delta^{A_1 A_2 B_2}_{D_1 C_2 D_2} \left[ \Kt^{D_1}_{A_1} \Kt^{D_2 \tilde{D_2}} \mathcal{R}^{C_2}_{A_2\tilde{D_2} B_2} +  \nabla_{A_2} \Kt^{D_1}_{A_1}\nabla^{D_2}\Kt^{C_2}_{B_2} \right] \,.
\end{split}
\ee
The first term in Eq.~\eqref{cont3app} is a total derivative. However, it plays a crucial role as an obstruction to  our analysis being complete and we can not ignore it. The remaining terms are described in the text following Eq.~\eqref{cont3}.

\section{Equations of motion in Lovelock theories} 
\label{app:loveDvv}

We present a quick derivation of the equations of motion in Lovelock theories described by the action Eq.~\eqref{eq:LoveAction}. Since the derivation is almost identical to the Gauss-Bonnet case (see \S\ref{appA3}), we will be very brief.

Consider the contribution to the gravitation equations of motion from a particular Lovelock term. By varying the action $m^{\rm th}$ Lovelock Lagrangian $\mathcal{L}_m$, we find the following:
\begin{equation}\label{Eom1L}
\begin{split}
\mathcal{E}_{\mu \nu}^{(m)} &= \frac{\delta\mathcal{L}_m }{\delta g^{\mu \nu }} - \frac{g_{\mu \nu }}{2} \mathcal{L}_m \\
&=m\,\delta^{\mu_1 \nu_1\cdots \mu_m \nu_m}_{\rho_1 \sigma_1\cdots\rho_m \sigma_m} \bigg[{ R^{\rho_1}}_{\mu_1(\mu\nu_1}\delta^{\sigma_1}_{\nu)} + g^{\alpha\sigma_1}\left(D_{\alpha} \frac{\delta\Gamma^{\rho_1}_{\mu_1 \nu_1}}{\delta g^{\mu \nu}} -D_{\nu_1} \frac{\delta\Gamma^{\rho_1}_{\mu_1 \alpha}}{\delta g^{\mu \nu}}\right) \bigg] \\
&\qquad \times {{{{R^{\rho_2}}_{\mu_2}}^{\sigma_2}}_{\nu_2}}\cdots {{{{R^{\rho_m}}_{\mu_m}}^{\sigma_m}}_{\nu_m}}  - \frac{g_{\mu \nu }}{2}  \mathcal{L}_m\,, \\
\end{split}
\end{equation}
where the factor $m$ comes from the variation of a product of $m$ Riemann tensors.
As before, the part involving variations of Christoffel symbols is zero (up to total derivative terms, which do not contribute):
\begin{equation}\label{Eom2L}
\begin{split}
&\delta^{\mu_1 \nu_1\cdots \mu_m \nu_m}_{\rho_1 \sigma_1\cdots\rho_m \sigma_m} ~g^{\alpha\sigma_1}\left(D_{\alpha} \delta\Gamma^{\rho_1}_{\mu_1 \nu_1} -D_{\nu_1} \Gamma^{\rho_1}_{\mu_1 \alpha}\right){{{{R^{\rho_2}}_{\mu_2}}^{\sigma_2}}_{\nu_2}}\cdots {{{{R^{\rho_m}}_{\mu_m}}^{\sigma_m}}_{\nu_m}} \\
&=(m-1)\delta^{\mu_1 \nu_1\cdots \mu_m \nu_m}_{\rho_1 \sigma_1\cdots\rho_m \sigma_m} \left[ D^{\sigma_1} \delta g^{\rho_1}_{\mu_1} \right](D_{\nu_1} {{{{R^{\rho_2}}_{\mu_2}}^{\sigma_2}}_{\nu_2}}){{{{R^{\rho_3}}_{\mu_3}}^{\sigma_3}}_{\nu_3}}\cdots {{{{R^{\rho_m}}_{\mu_m}}^{\sigma_m}}_{\nu_m}} \\
&= 0 \,.
\end{split}
\end{equation}
where we used the Bianchi identity, Eq.~\eqref{bianchi1}.
Therefore the equations of motion take the form
\begin{equation}\label{Eom3L}
\begin{split}
(\mathcal{E}^{(m)})^{\alpha}_{\beta} &= m\, \delta^{\mu_1 \nu_1\cdots \mu_m \nu_m}_{\rho_1 \beta\cdots\rho_m \sigma_m}  {{{{R^{\rho_1}}_{\mu_1}}^{\alpha}}_{\nu_1}} {{{{R^{\rho_2}}_{\mu_2}}^{\sigma_2}}_{\nu_2}} \cdots {{{{R^{\rho_m}}_{\mu_m}}^{\sigma_m}}_{\nu_m}}
 - \frac{\delta^{\alpha}_{\beta}}{2}  \mathcal{L}_m\\
 &= \left( m\, \delta^\alpha_{{\sigma}_1} \delta_\beta^{\tilde{\sigma}_1} \,\delta^{\mu_1 \nu_1\cdots \mu_m \nu_m}_{\rho_1 \tilde{\sigma}_1\cdots\rho_m \sigma_m}  - \frac{1}{2} \;  \delta^{\alpha}_{\beta} \,\delta^{\mu_1\nu_1\cdots \mu_m \nu_m}_{\rho_1\sigma_1\cdots \rho_m \sigma_m}  \right) {{{{R^{\rho_1}}_{\mu_1}}^{{\sigma}_1}}_{\nu_1}} \cdots {{{{R^{\rho_m}}_{\mu_m}}^{\sigma_m}}_{\nu_m}} \\
 &=-\frac{1}{2}\delta^{\alpha \mu_1 \nu_1 \cdots \mu_m \nu_m}_{\beta \rho_1 \sigma_1\cdots \rho_m \sigma_m}\left[{{{{R^{\rho_1}}_{\mu_1}}^{\sigma_1}}_{\nu_1}} \cdots {{{{R^{\rho_m}}_{\mu_m}}^{\sigma_m}}_{\nu_m}}\right]\,.
  \end{split}
 \end{equation}
 From here, one can evaluate the temporal component $\mathcal{E}_{vv}^{(m)}$ to obtain Eq.~\eqref{eq:love1}.

\section{Analysis for $f(R)$ theory}
\label{sec:fR}

In this appendix, we review the proof of second law for $f(R)$ theories  given by \cite{Jacobson:1995uq} in our notation. 
In Gauss-Bonnet and Lovelock theory  to prove that $\Theta\geq0$, what we have attempted to show is the following
\begin{equation}\label{ineq}
\begin{split}
&\partial_v\Theta + \mathcal{E}_{vv}\leq 0\\
&\text{where the equation of motion:}~~ \mathcal{E}_{\mu\nu} -T_{\mu\nu} =0
\end{split}
\end{equation}
However  for our purpose proving the following inequality would have been enough
\begin{equation}\label{ineq2}
\partial_v(Z_1\Theta) + Z_2 \mathcal{E}_{vv}\leq 0 \,,
\end{equation}
where $Z_1$ and $Z_2$ are any two functions that are strictly positive within the validity regime of our approximation.
The proof of `Wald entropy increase'  in  \cite{Jacobson:1995uq} could be recast in this language with a certain choice of $Z_1$ and $Z_2$.

For the case of $f(R)$ theories, the Lagrangian is of the form
\begin{equation}\label{actionRsq}
I = \frac{1}{4\pi} \int  d^dx \sqrt{-g} \;f(R)
\end{equation}
From this, it can be shown that, see \cite{Jacobson:1995uq},\begin{equation}\label{S3}
\begin{split}
& S_\text{Wald} = \int_{\Sigma_v} d^{d-2}x\,\sqrt{h}~f'(R) \,.
\end{split}
\end{equation}
By taking one more derivative, we find
\begin{equation}\label{Theta3}
\begin{split}
\Theta &= \frac{1}{\sqrt{h}} \partial_v\left[\sqrt{h}f'(R)\right] = \Kt\, f'(R)+\partial_vf'(R) \,.
\end{split}
\end{equation}
In such theories the $(vv)$ component of the equation of motion takes the following form:
\begin{equation}\label{EOM}
\begin{split}
\mathcal{E}_{vv} =&~ f'(R)R_{vv} - \partial_v^2 f'(R)\\
=&~ f'(R)\left[R_{vv} - \frac{\partial_v^2 f'(R)}{f'(R)}\right]\\
=&~ f'(R)\left[R_{vv} - \partial^2_v \log[f'(R)]-\left(\frac{\partial_v f'(R)}{f'(R)}\right)^2\right]\,.
\end{split}
\end{equation}
Now define $Z_1 =Z_2 = \frac{1}{f'(R)}$
and substitute in the LHS of equation Eq.~\eqref{ineq2}:
\begin{equation}\label{z1z2}
\begin{split}
&\partial_v\left(\frac{\Theta}{f'(R)}\right) + \frac{\mathcal{E}_{vv}}{f'(R)}\\
=~&\partial_v \Kt +\partial_v^2\log[f'(R)] +\left[R_{vv} - \partial^2_v \log[f'(R)]-\left(\frac{\partial_v f'(R)}{f'(R)}\right)^2\right]\\
=~&R_{vv} +\partial_v \Kt -\left(\frac{\partial_v f'(R)}{f'(R)}\right)^2\\
=~&-\Kt_{AB} \Kt^{AB} -\left(\frac{\partial_v f'(R)}{f'(R)}\right)^2 \leq 0 \,.
\end{split}
\end{equation}
Here in the second line we have used equations Eq.~\eqref{Theta3} and Eq.~\eqref{EOM}. In the final step we have used the fact that
$$R_{vv}\bigg \vert_{Horizon} = -\partial_v \Kt - \Kt_{AB} \Kt^{AB}$$

Now in our case, $f'(R) = 1 + {\cal O}(\omega^2\,\ell_s^2)$.
Hence both $Z_1$ and $Z_2$ are strictly positive.



\begin{thebibliography}{10}

\bibitem{Adams:2006sv}
A.~Adams, N.~Arkani-Hamed, S.~Dubovsky, A.~Nicolis, and R.~Rattazzi, {\it
  {Causality, analyticity and an IR obstruction to UV completion}},  {\em JHEP}
  {\bf 10} (2006) 014, [\href{http://arxiv.org/abs/hep-th/0602178}{{\tt
  hep-th/0602178}}].

\bibitem{Camanho:2014apa}
X.~O. Camanho, J.~D. Edelstein, J.~Maldacena, and A.~Zhiboedov, {\it {Causality
  Constraints on Corrections to the Graviton Three-Point Coupling}},  {\em
  JHEP} {\bf 02} (2016) 020, [\href{http://arxiv.org/abs/1407.5597}{{\tt
  arXiv:1407.5597}}].

\bibitem{Hawking:1971tu}
S.~W. Hawking, {\it {Gravitational radiation from colliding black holes}},
  {\em Phys. Rev. Lett.} {\bf 26} (1971) 1344--1346.

\bibitem{Bardeen:1973gs}
J.~M. Bardeen, B.~Carter, and S.~W. Hawking, {\it {The Four laws of black hole
  mechanics}},  {\em Commun. Math. Phys.} {\bf 31} (1973) 161--170.

\bibitem{Bekenstein:1973ur}
J.~D. Bekenstein, {\it {Black holes and entropy}},  {\em Phys. Rev.} {\bf D7}
  (1973) 2333--2346.

\bibitem{Hawking:1974sw}
S.~W. Hawking, {\it {Particle Creation by Black Holes}},  {\em Commun. Math.
  Phys.} {\bf 43} (1975) 199--220. [,167(1975)].

\bibitem{Hawking:1973uf}
S.~W. Hawking and G.~F.~R. Ellis, {\em {The Large Scale Structure of
  Space-Time}}.
\newblock Cambridge Monographs on Mathematical Physics. Cambridge University
  Press, 2011.

\bibitem{Wald:1993nt}
R.~M. Wald, {\it {Black hole entropy is the Noether charge}},  {\em Phys. Rev.}
  {\bf D48} (1993), no.~8 R3427--R3431,
  [\href{http://arxiv.org/abs/gr-qc/9307038}{{\tt gr-qc/9307038}}].

\bibitem{Iyer:1994ys}
V.~Iyer and R.~M. Wald, {\it {Some properties of Noether charge and a proposal
  for dynamical black hole entropy}},  {\em Phys. Rev.} {\bf D50} (1994)
  846--864, [\href{http://arxiv.org/abs/gr-qc/9403028}{{\tt gr-qc/9403028}}].

\bibitem{Jacobson:1993xs}
T.~Jacobson and R.~C. Myers, {\it {Black hole entropy and higher curvature
  interactions}},  {\em Phys. Rev. Lett.} {\bf 70} (1993) 3684--3687,
  [\href{http://arxiv.org/abs/hep-th/9305016}{{\tt hep-th/9305016}}].

\bibitem{Jacobson:1993vj}
T.~Jacobson, G.~Kang, and R.~C. Myers, {\it {On black hole entropy}},  {\em
  Phys. Rev.} {\bf D49} (1994) 6587--6598,
  [\href{http://arxiv.org/abs/gr-qc/9312023}{{\tt gr-qc/9312023}}].

\bibitem{Jacobson:1995uq}
T.~Jacobson, G.~Kang, and R.~C. Myers, {\it {Increase of black hole entropy in
  higher curvature gravity}},  {\em Phys. Rev.} {\bf D52} (1995) 3518--3528,
  [\href{http://arxiv.org/abs/gr-qc/9503020}{{\tt gr-qc/9503020}}].

\bibitem{Kolekar:2012tq}
S.~Kolekar, T.~Padmanabhan, and S.~Sarkar, {\it {Entropy Increase during
  Physical Processes for Black Holes in Lanczos-Lovelock Gravity}},  {\em Phys.
  Rev.} {\bf D86} (2012) 021501, [\href{http://arxiv.org/abs/1201.2947}{{\tt
  arXiv:1201.2947}}].

\bibitem{Sarkar:2013swa}
S.~Sarkar and A.~C. Wall, {\it {Generalized second law at linear order for
  actions that are functions of Lovelock densities}},  {\em Phys. Rev.} {\bf
  D88} (2013) 044017, [\href{http://arxiv.org/abs/1306.1623}{{\tt
  arXiv:1306.1623}}].

\bibitem{Bhattacharjee:2015yaa}
S.~Bhattacharjee, S.~Sarkar, and A.~C. Wall, {\it {Holographic entropy
  increases in quadratic curvature gravity}},  {\em Phys. Rev.} {\bf D92}
  (2015), no.~6 064006, [\href{http://arxiv.org/abs/1504.04706}{{\tt
  arXiv:1504.04706}}].

\bibitem{Wall:2015raa}
A.~C. Wall, {\it {A Second Law for Higher Curvature Gravity}},  {\em Int. J.
  Mod. Phys.} {\bf D24} (2015), no.~12 1544014,
  [\href{http://arxiv.org/abs/1504.08040}{{\tt arXiv:1504.08040}}].

\bibitem{Bhattacharjee:2015qaa}
S.~Bhattacharjee, A.~Bhattacharyya, S.~Sarkar, and A.~Sinha, {\it {Entropy
  functionals and c-theorems from the second law}},  {\em Phys. Rev.} {\bf D93}
  (2016), no.~10 104045, [\href{http://arxiv.org/abs/1508.01658}{{\tt
  arXiv:1508.01658}}].

\bibitem{Fursaev:2013fta}
D.~V. Fursaev, A.~Patrushev, and S.~N. Solodukhin, {\it {Distributional
  Geometry of Squashed Cones}},  {\em Phys. Rev.} {\bf D88} (2013), no.~4
  044054, [\href{http://arxiv.org/abs/1306.4000}{{\tt arXiv:1306.4000}}].

\bibitem{Dong:2013qoa}
X.~Dong, {\it {Holographic Entanglement Entropy for General Higher Derivative
  Gravity}},  {\em JHEP} {\bf 01} (2014) 044,
  [\href{http://arxiv.org/abs/1310.5713}{{\tt arXiv:1310.5713}}].

\bibitem{Camps:2013zua}
J.~Camps, {\it {Generalized entropy and higher derivative Gravity}},  {\em
  JHEP} {\bf 03} (2014) 070, [\href{http://arxiv.org/abs/1310.6659}{{\tt
  arXiv:1310.6659}}].

\bibitem{Bhattacharyya:2008xc}
S.~Bhattacharyya, V.~E. Hubeny, R.~Loganayagam, G.~Mandal, S.~Minwalla,
  T.~Morita, M.~Rangamani, and H.~S. Reall, {\it {Local Fluid Dynamical Entropy
  from Gravity}},  {\em JHEP} {\bf 06} (2008) 055,
  [\href{http://arxiv.org/abs/0803.2526}{{\tt arXiv:0803.2526}}].

\bibitem{Bhattacharyya:2008jc}
S.~Bhattacharyya, V.~E. Hubeny, S.~Minwalla, and M.~Rangamani, {\it {Nonlinear
  Fluid Dynamics from Gravity}},  {\em JHEP} {\bf 02} (2008) 045,
  [\href{http://arxiv.org/abs/0712.2456}{{\tt arXiv:0712.2456}}].

\bibitem{Hubeny:2011hd}
V.~E. Hubeny, S.~Minwalla, and M.~Rangamani, {\it {The fluid/gravity
  correspondence}},  in {\em {Black holes in higher dimensions}}, pp.~348--383,
  2012.
\newblock \href{http://arxiv.org/abs/1107.5780}{{\tt arXiv:1107.5780}}.
\newblock [,817(2011)].

\bibitem{Brigante:2007nu}
M.~Brigante, H.~Liu, R.~C. Myers, S.~Shenker, and S.~Yaida, {\it {Viscosity
  Bound Violation in Higher Derivative Gravity}},  {\em Phys. Rev.} {\bf D77}
  (2008) 126006, [\href{http://arxiv.org/abs/0712.0805}{{\tt
  arXiv:0712.0805}}].

\bibitem{Marolf:2013ioa}
D.~Marolf, M.~Rangamani, and T.~Wiseman, {\it {Holographic thermal field theory
  on curved spacetimes}},  {\em Class. Quant. Grav.} {\bf 31} (2014) 063001,
  [\href{http://arxiv.org/abs/1312.0612}{{\tt arXiv:1312.0612}}].

\bibitem{wald2010general}
R.~Wald, {\em General Relativity}.
\newblock University of Chicago Press, 2010.

\bibitem{Bhattacharyya:2013lha}
S.~Bhattacharyya, {\it {Entropy current and equilibrium partition function in
  fluid dynamics}},  {\em JHEP} {\bf 08} (2014) 165,
  [\href{http://arxiv.org/abs/1312.0220}{{\tt arXiv:1312.0220}}].

\bibitem{Bhattacharyya:2014bha}
S.~Bhattacharyya, {\it {Entropy Current from Partition Function: One Example}},
   {\em JHEP} {\bf 07} (2014) 139, [\href{http://arxiv.org/abs/1403.7639}{{\tt
  arXiv:1403.7639}}].

\bibitem{Haehl:2015pja}
F.~M. Haehl, R.~Loganayagam, and M.~Rangamani, {\it {Adiabatic hydrodynamics:
  The eightfold way to dissipation}},  {\em JHEP} {\bf 05} (2015) 060,
  [\href{http://arxiv.org/abs/1502.00636}{{\tt arXiv:1502.00636}}].

\bibitem{Oliva:2010eb}
J.~Oliva and S.~Ray, {\it {A new cubic theory of gravity in five dimensions:
  Black hole, Birkhoff's theorem and C-function}},  {\em Class. Quant. Grav.}
  {\bf 27} (2010) 225002, [\href{http://arxiv.org/abs/1003.4773}{{\tt
  arXiv:1003.4773}}].

\bibitem{Myers:2010ru}
R.~C. Myers and B.~Robinson, {\it {Black Holes in Quasi-topological Gravity}},
  {\em JHEP} {\bf 08} (2010) 067, [\href{http://arxiv.org/abs/1003.5357}{{\tt
  arXiv:1003.5357}}].

\bibitem{Oliva:2011xu}
J.~Oliva and S.~Ray, {\it {Birkhoff's Theorem in Higher Derivative Theories of
  Gravity}},  {\em Class. Quant. Grav.} {\bf 28} (2011) 175007,
  [\href{http://arxiv.org/abs/1104.1205}{{\tt arXiv:1104.1205}}].

\bibitem{Deser:2016tgn}
S.~Deser, {\it {One-loop gravity divergences in D>4 cannot all be removed}},
  {\em Gen. Rel. Grav.} {\bf 48} (2016), no.~12 157,
  [\href{http://arxiv.org/abs/1609.04432}{{\tt arXiv:1609.04432}}].

\bibitem{Bhattacharyya:2016nhn}
S.~Bhattacharyya, A.~K. Mandal, M.~Mandlik, U.~Mehta, S.~Minwalla, U.~Sharma,
  and S.~Thakur, {\it {Currents and Radiation from the large $D$ Black Hole
  Membrane}},  \href{http://arxiv.org/abs/1611.09310}{{\tt arXiv:1611.09310}}.

\end{thebibliography}

\providecommand{\href}[2]{#2}\begingroup\raggedright\endgroup

\end{document}